astro-ph/9311073 29 Nov 93

# Geometrical Evidence for Dark Matter: X-ray Constraints on the Mass of the Elliptical Galaxy NGC 720

David A. Buote[1] and Claude R. Canizares[2]

Department of Physics and Center for Space Research 37-241,

Massachusetts Institute of Technology

77 Massachusetts Avenue, Cambridge, MA 02139

## ABSTRACT

We describe (1) a new test for dark matter and alternate theories of gravitation based on the relative geometries of the X-ray and optical surface brightness distributions and an assumed form for the potential of the optical light, (2) a technique to measure the shapes of the total gravitating matter and dark matter of an ellipsoidal system which is insensitive to the precise value of the temperature of the gas and to modest temperature gradients, and (3) a method to determine the ratio of dark mass to stellar mass that is dependent on the functional forms for the visible star, gas and dark matter distributions, but independent of the distance to the galaxy or the gas temperature.

We apply these techniques to X-ray data from the ROSAT Position Sensitive Proportional Counter (PSPC) of the optically-flattened elliptical galaxy NGC 720; the optical isophotes have ellipticity $\epsilon \sim 0.40$ extending out to $\sim 120''$ ($10'' \sim 1$ kpc assuming a distance of $20h_{80}$ Mpc). The X-ray isophotes are significantly elongated, $\epsilon = 0.20 - 0.30$ (90% confidence) for semi-major axis $a \sim 100''$. The major axes of the optical and X-ray isophotes are misaligned by $\sim 30° \pm 15°$ (90% confidence). Spectral analysis of the X-ray data reveals no evidence of temperature gradients or anisotropies and demonstrates that a single-temperature plasma ($T \sim 0.6$ keV) having sub-solar heavy element abundances and a two-temperature model having solar abundances describe the spectrum equally well. Considering only the relative geometries of the X-ray and optical surface brightness distributions and an assumed functional form for the potential of the optical light, *we conclude that matter distributed like the optical light cannot produce the observed ellipticities of the X-ray isophotes*, independent of the gas pressure, the gas temperature, and the value of the stellar mass; this comparison assumes a state of quasi-hydrostatic equilibrium so that the three-dimensional surfaces of constant gas emissivity trace the three-dimensional isopotential surfaces – we discuss the viability of this assumption for NGC 720. Milgrom's Modification of Newtonian Dynamics (MOND) cannot dispel this manifestation of dark matter. Hence, geometrical considerations, which are essentially independent of gas pressure or temperature, require the presence of an extended, massive dark matter halo in NGC 720.

---

[1] dbuote@space.mit.edu

[2] crc@space.mit.edu

Employing essentially the technique of Buote & Canizares (1992; Buote 1992) we use the *shape* of the X-ray surface brightness to constrain the *shape* of the total gravitating matter. The total matter is modeled as either an oblate or prolate spheroid of constant shape and orientation having either a Ferrers ($\rho \sim r^{-n}$) or Hernquist density. Assuming the X-ray gas is in hydrostatic equilibrium, we construct a model X-ray gas distribution for various temperature profiles; i.e. isothermal, linear, and polytropic. We determine the ellipticity of the total gravitating matter to be $\epsilon \sim 0.50 - 0.70$. Using the single-temperature model we estimate a total mass $\sim (0.41 - 1.4) \times 10^{12} h_{80} M_{\odot}$ interior to the ellipsoid of semi-major axis $43.6 h_{80}$ kpc. Ferrers densities as steep as $r^{-3}$ do not fit the data, but the $r^{-2}$ and Hernquist models yield excellent fits. We estimate the mass distributions of the stars and the gas and fit the dark matter directly. For a given temperature profile of the gas and functional forms for the visible stars, gas, and dark matter, these models yield a distance-independent and temperature-independent measurement of the ratio of dark mass to stellar mass $M_{DM}/M_{stars}$. We estimate at minimum $M_{DM}/M_{stars} \geq 4$ which corresponds to a total mass slightly greater than that derived from the single-temperature models for distance $D = 20 h_{80}$ Mpc.

*Subject headings:* dark matter — galaxies: elliptical and lenticular, cD — galaxies: halos — galaxies: individual (NGC 720) — galaxies: kinematics and dynamics — galaxies: photometry — galaxies: structure — galaxies: X-rays — gravitation — interstellar medium: structure

# 1. Introduction

The nature and distribution of dark matter in the universe persists as one of the most important unresolved problems in astrophysics. Although preciously little is known about the nature of the dark matter, strong constraints on its radial distribution exist on galactic scales from the flat H I rotation curves in spiral galaxies (for reviews see Kormendy & Knapp 1987; Trimble 1987; Ashman 1992) and, recently, gravitational lens models of luminous arcs in clusters of galaxies (for reviews see Blandford & Narayan 1992; Soucail 1992; Refsdal & Surdej 1993) However, there is comparatively little convincing evidence for dark matter in normal elliptical galaxies (for reviews see Kent 1990; de Zeeuw & Franx 1991; Ashman 1992); this lack of evidence is generally attributed to the fact that most optical studies are confined to within an effective radius of the galaxy center where the effects of dark matter may be unimportant.

X-ray emission from hot gas provides perhaps the greatest potential for accurately mapping the mass of ellipticals to large distances (for a review see Fabbiano 1989). The standard method employed to infer the mass from the X-ray gas derives from the equation of hydrostatic equilibrium and the ideal gas equation of state (Fabricant, Lecar, & Gorenstein 1980),

$$M(<r) = -\frac{r k_B T_{gas}(r)}{G \mu m_p} \left( \frac{d \ln \rho_{gas}}{d \ln r} + \frac{d \ln T_{gas}}{d \ln r} \right),$$ (1)

where $T_{gas}$ is the gas temperature, $\rho_{gas}$ is the gas density, $G$ is Newton's constant, $k_B$ is Boltzmann's constant, $\mu$ is the mean atomic weight of the gas, and $m_p$ is the proton mass; note that this method assumes spherical symmetry of the mass distribution. At a given $r$, equation (1) has three quantities to be determined from observations; i.e. the gas temperature, temperature gradient, and density gradient. Unfortunately, attempts to apply this technique to *Einstein* data of normal elliptical galaxies yielded very uncertain results because the normal ellipticals had poorly determined temperature profiles. For example, Trinchieri, Fabbiano, & Canizares (1986) analyzed *Einstein* images of six early-type galaxies and concluded that the X-ray data were consistent with massive dark halos but halos were not absolutely required by the data. Similarly, employing the improved spectral resolution of the BBXRT to the Virgo elliptical NGC 4472, Serlemitsos et al. (1993), determine that the data does not demand dark matter. However, they conclude that the BBXRT data for the Fornax elliptical NGC 1399 indeed requires significant amounts of dark matter.

Whereas the previous technique embodied by equation (1) probes the radial mass distribution, White (1987; White & Canizares 1987), who built upon the pioneering study of Binney & Strimple (1978; Strimple & Binney 1979), introduced a modification of this method to measure the shape of the total gravitating matter in their study of the elliptical galaxy NGC 720 as well as two other early-type galaxies, NGC 1332 and NGC 4697. They relax the assumption of spherical symmetry and assume the gas is isothermal. By using the ratio of potential depth to gas temperature as a fitting parameter, their method is very insensitive to the precise value of the gas temperature. However, they were still unable to obtain meaningful constraints on the shape of the underlying matter because of the large point spread function of the *Einstein* Imaging Proportional Counter (IPC). Buote & Canizares (1992; Buote 1992) utilized the technique of White (1987; White & Canizares 1987) for analysis of five Abell clusters of galaxies. Because of the larger IPC fluxes and spatial extent of the clusters, Buote & Canizares succeeded in measuring the shape of the gravitating matter; they determined the total matter to be significantly rounder than the galaxy isopleths for all of the clusters.

We improve upon the technique of Buote & Canizares (1992; Buote 1992) to measure the shape and amount of dark matter in the flattened elliptical NGC 720 using the superior X-ray data provided by the *Röntgen Satellite* (ROSAT). By assuming functional forms for the mass of the visible stars, X-ray gas, and dark matter, our method enables direct measurement of the shape of not only the total gravitating matter, but also the dark matter itself; we show that this method yields a mass estimate that is independent of the distance to the galaxy and the temperature of the gas. In addition, by exploiting the relative geometries of the X-ray and optical isophotes (and an assumed model for the potential of the optical light) we introduce a test for dark matter and alternate gravity theories that is highly insensitive to uncertainties in the gas temperature. In §2. we discuss the observations and determination of the relevant parameters required for the analysis; in §3. we describe our geometrical test for dark matter; in §4. we measure the shape and amount of total gravitating matter; in §5. we do the same for the dark matter, in §6. we discuss the implications of our results; and in §7. we present our conclusions.

## 2. Observations and Data Analysis

We selected NGC 720 for analysis as one of two early-type galaxies with flattened optical morphology and high X-ray flux as measured by *Einstein* (e.g., Fabbiano, Kim, & Trinchieri 1992). The optical isophotes have ellipticity $\epsilon \sim 0.40$, which makes NGC 720 one of the flattest ellipticals, and suggests that its intrinsic shape is close to its projection on the sky (Fasano & Vio 1991; Ryden 1991, 1992; Lambas, Maddox, & Loveday 1992). Here $\epsilon$ is defined as $1 - b/a$ where $a$ ($b$) is the major (minor) axis. Assuming the elongation of the stellar distribution indicates elongation of any putative dark matter, then one would expect the X-ray isophotes tracing the gravitational potential (which, however, is always rounder than the parent mass) would be most elongated for galaxies with the flattest optical isophotes. NGC 720 possesses the largest X-ray flux (e.g., Fabbiano et al. 1992) for a flattened normal galaxy and its emission extends over 13′ on the sky, thus providing many pixels of angular resolution. In addition, NGC 720 is a relatively isolated elliptical (Dressler, Schechter, & Rose 1986) suggesting that its X-ray emission is mostly free of contamination from external effects such as ram-pressure stripping (Schechter 1987). The galaxy was observed with the Position Sensitive Proportional Counter (PSPC) on board ROSAT; for a description of the ROSAT X-ray Telescope see Aschenbach (1988), and Pfeffermann et al. (1987) for a description of the PSPC. Table 1 summarizes the details of the observation.

The distance to NGC 720 has been determined by several different methods, including Hubble flow analysis (e.g., in Canizares, Fabbiano, & Trinchieri 1987), $D_n - \sigma$ (Donnely, Faber, & O'Connell 1990), and surface brightness fluctuations (Tonry & Blakeslee 1993, private communication). The values derived from these methods systematically differ by as much as 9 Mpc with the $D_n - \sigma$ estimate representing the high end (24.8$h_{80}$ Mpc) and the surface brightness fluctuations the lower end (15.6$h_{80}$ Mpc). We adopt $D = 20h_{80}$ Mpc as essentially a mean value for the distance to NGC 720 where $h_{80} \equiv 1$ for $H_0 = 80$ km s$^{-1}$ Mpc$^{-1}$; at this distance $1'' \sim 0.1$ kpc.

### 2.1. Spatial Analysis

We rebinned the PSPC image of NGC 720 into 15″ pixels, corresponding to a $512 \times 512$ field, which effectively optimizes the signal-to-noise ratio. In order to minimize the X-ray background contribution to the galaxy emission and to optimize the PSF of the PSPC (see below), only data from the hard band (0.4 - 2.4 keV) were used.

Employing the standard IRAF-PROS software, we constructed a surface brightness map from the observation by (1) correcting for exposure variations and telescopic vignetting, (2) removing embedded point sources, and (3) subtracting the background. The vignetting correction for NGC 720 is small since only a few percent of the total emission from the galaxy lies $> 6'$ off-axis where this effect becomes important. The standard processing routines identify point sources in the field using a maximum-likelihood method (cf. *detect* task in PROS). We used this source list as a guide to flag sources not associated with the continuum emission of the galaxy. Three additional point sources not included in this list were identified "by eye". All of these sources were flagged and excluded from succeeding analysis.

The final step in the image reduction is the estimation and subtraction of the background. The in-flight software identifies and eliminates effects of the particle background (Snowden et al. 1992). For the remainder of the X-ray background, the standard processing of the observation generates a template to serve as a convenient background estimate. These templates are constructed by subtracting all of the point sources out of the image and then smoothing. For sources with extended emission, these templates may overestimate the background due to incomplete subtraction of the extended source. We investigated this effect by binning the image (corrected as above) into $15''$ radial bins centered on the galaxy emission (cf. §2.1.1.). In Figure 1 we plot the azimuthally-averaged radial profile of the image between $100''$ and $1000''$ and compare to the corresponding region of the background template; we do not assign error bars to the background because the systematic errors dominate any statistical errors as a result of the heavy processing of the template. The statistical errors assigned to the image are 68% Poisson confidence limits obtained using the approximate expressions of Gehrels (1986). The template matches the image to better than a few percent for radii greater than $400''$ where the background should dominate the galaxy emission; although the figure displays a slight rise in the background towards the center, for $r \lesssim 100''$ small errors in the background are unimportant since there the galaxy emission dominates. Hence, the template represents the background adequately for our purposes. We subtract the background template from the image and use only the statistical uncertainties associated with the image in our analysis.

Figure 2 displays isophotes for the reduced image in a $400'' \times 400''$ region centered on the galaxy. We confine our analysis to the region interior to $375''$ as that is where the signal-to-noise (S/N) in each bin is $\gtrsim 1$. We have smoothed the image in Figure2 with a circular Gaussian ($\sigma = 11.25''$) for visual clarity although we emphasize that the image used for analysis is not smoothed in this manner.

### 2.1.1. Radial Profile

We constructed the azimuthally-averaged radial profile of the X-ray surface brightness as follows. We located the origin of the radial profile at the centroid of the galaxy emission $(\bar{x}, \bar{y})$ determined by computing the first moments of the count distribution,

$$\bar{x} = \frac{1}{N} \sum_{i=1}^{P} n_i x_i \quad \text{and} \quad \bar{y} = \frac{1}{N} \sum_{i=1}^{P} n_i y_i, \tag{2}$$

where $i$ denotes the label of the pixel, $P$ represents the total number of pixels included in the summation, $n_i$ is the number of counts in pixel $i$, $(x_i, y_i)$ are the Cartesian coordinates of pixel $i$, and $N = \sum_{i=1}^{P} n_i$ is the total number of counts in pixels $P$. After choosing a center of the galaxy counts "by eye", the moments were computed within a $150''$ circular aperture containing $\sim 75\%$ of the total counts ($< 375''$) and then iterated until the centroid varied by $< 0.1\%$. The centroid position obtained, listed in Table 1, agrees with the optical position to a fraction of a pixel. Next we binned the counts into circular annuli of one pixel width (i.e. $15''$) centered at $(\bar{x}, \bar{y})$; we explored the effect of using elliptical annuli having the shapes and orientations of the isophotes (cf. 2.1.2.) but found no appreciable gain in S/N. The radial profile of the reduced image is displayed in Figure 3. The shape of the radial profile is not particularly sensitive to the initial guess of the centroid or to the size of the centroid aperture.

Previous studies of the X-ray surface brightness distribution $(\Sigma_X)$ of galaxies (e.g., Forman, Jones, & Tucker 1985; Trinchieri, Fabbiano, & Canizares 1986; hereafter TFC) used the hydrostatic-isothermal King-type model to parameterize $\Sigma_X$,

$$\Sigma_X(r) \propto \left[ 1 + \left( \frac{r}{a_X} \right)^2 \right]^{-3\beta + 1/2}, \tag{3}$$

where $a_X$ and $\beta$ are free parameters. The assumption of spherical symmetry in the King model, although not strictly valid for the galaxy isophotes (cf. § 2.1.2.), has a small effect on fits to the surface brightness profile of NGC 720. The King model serves as a convenient analytic fit to $\Sigma_X$, which facilitates computation of the mass of the X-ray gas (§5.2.). In order to obtain physical constraints on $a_X$ and $\beta$, we convolve $\Sigma_X$ with the PSPC PSF and perform a $\chi^2$ fit to the radial profile. The on-axis PSF described by Hasinger et al. (1992) depends on the energy of the incident photon and is composed of a circular Gaussian component due to intrinsic broadening, an exponential component due to focus and photon penetration effects, and a Lorentzian component due to mirror scattering. Performing a counts-weighted average of the galaxy spectrum between 0.4 and 2.4 keV, we adopt $E = 0.88$ keV for evaluation of the PSF. We list the results of the fit in Table 2 along with those published by TFC; the best fit model is plotted in Figure 3. The values for $a_X$ obtained by TFC for NGC 720 with the *Einstein* Imaging Proportional Counter (IPC) agree very well with our values, essentially bracketing our results. However, TFC's 90% confidence limits for the slope parameter $\beta$ are slightly smaller. We can attribute this difference to the fact that TFC include emission from point sources that we have identified and eliminated; the number of such sources increases with distance from the galaxy center. This effect will tend to flatten their radial profile sufficiently to account for the slight systematic shift in $\beta$.

Note that the 90% lower limit for $a_x = 12''$ is significantly larger than the optical core radius of $4''$ (§5.1.). If the temperature gradients are small (which we show in §2.2.), then the hydrostatic equation (eq. [1]) implies that the total matter must have a core parameter similar in magnitude to $a_X$. Hence, the discrepancy between X-ray and optical core parameters suggests that the total mass can not be described simply by matter distributed like the visible stars. We address this issue in more detail in later sections.

### 2.1.2. Ellipticities of the X-ray Isophotes

In comparison to optical images of ellipticals, the PSPC X-ray image of NGC 720 has significantly fewer counts ($\sim 1500$ for $r < 200''$). As a result, our analysis of the morphology of the X-ray surface brightness more closely parallels the analysis of the galaxy isopleths in a rich cluster than the optical isophotes of an elliptical. We measure the flattening and orientation of the X-ray surface brightness using an iterative moment technique derived from the treatment of the dispersion ellipse of the bivariate normal frequency function of position vectors used by Carter & Metcalf (1980; Trumpler & Weaver 1953) to measure the ellipticities of clusters of galaxies. The parameters obtained from this method, $\epsilon_M$ and $\theta_M$, computed within an elliptical region, provide good estimates of the ellipticity ($\epsilon$) and the position angle ($\theta$) of an intrinsic elliptical distribution of constant shape and orientation. For a more complex distribution, $\epsilon_M$ and $\theta_M$ are average values weighted heavily by the outer parts of the region; in Buote & Canizares (1992) we apply a slight variation of this method to the study of five Abell clusters.

In order to determine these parameters from an image of P pixels having $n_i$ counts in pixel $i$, one computes the moments,

$$\mu_{mn} = \frac{1}{N} \sum_{i=1}^{P} n_i (x_i - \bar{x})^m (y_i - \bar{y})^n \qquad (m, n \leq 2), \tag{4}$$

where as before $N = \sum_{i=1}^{P} n_i$, and $(\bar{x}, \bar{y})$ is the centroid given by equation (2). Then the ellipticity is,

$$\epsilon_M = 1 - \frac{\Lambda_-}{\Lambda_+}, \tag{5}$$

and the position angle of the major axis measured North through East in Celestial coordinates is,

$$\theta_M = \tan^{-1}\left( \frac{\mu_{11}}{\Lambda_+^2 - \mu_{02}} \right) + \frac{\pi}{2}, \tag{6}$$

where $\Lambda_{\pm} (\Lambda_+ \geq \Lambda_-)$ are the positive roots of the quadratic,

$$(\mu_{20} - \Lambda^2)(\mu_{02} - \Lambda^2) = \mu_{11}^2; \tag{7}$$

for an elliptical Gaussian distribution, $\Lambda_+$ and $\Lambda_-$ are the respective lengths of the semi-major and semi-minor axes of the contour representing 0.61 times the maximum surface density.

The assumption of a Gaussian distribution is not necessary since $\mu_{mn}$ is equivalent to the two-dimensional moment of inertia tensor with $\Lambda_+^2$ and $\Lambda_-^2$ its principle moments. For any elliptical distribution the square root of the ratio of the principle moments of inertia is the axial ratio and thus $\epsilon_M$ is the ellipticity.

We begin by defining a circular aperture ($\epsilon_M = 0$) about the centroid determined in §2.1.1. with the initial value of $\theta_M$ set arbitrarily to 0. Then we compute the appropriate $\mu_{mn}$ for all of the pixels in this aperture to obtain new values of $\epsilon_M$, $\theta_M$, $\bar{x}$, and $\bar{y}$. Defining a new elliptical aperture with these parameters, we iterate until the parameters change by less than appropriate tolerances. Using the same iterative procedure, we also compute $\epsilon_M$ and $\theta_M$ within an elliptical annular aperture. The annulus should correspond more closely to a true isophote since only counts in the immediate vicinity of the isophote are used. However, as we discuss below, the values of $\epsilon_M$ and $\theta_M$ do not significantly differ for the elliptical and annular apertures.

Characterization of the uncertainties in this procedure involves both statistical and systematic effects. Random uncertainties due to Poisson statistics are straightforward and we derive expressions for the 90% confidence estimates $\Delta \epsilon_M$ and $\Delta \theta_M$ in Appendix A.. Quantification of the systematic uncertainties associated with the computation of $\epsilon_M$ is more subtle and requires numerical simulations. Using simulated images, Carter & Metcalfe (1980) concluded that $\epsilon_M$ deviates from $\epsilon$ (true ellipticity) due to the following systematic effects:

1. For distributions where $\epsilon$ is small or zero, any random deviations will increase the measured value of $\epsilon_M$.

2. For distributions where $\epsilon$ is large, the initial ellipticity of the circular aperture is far from the desired value. The iteration can get caught in a local stable point at a small value of $\epsilon_M$.

The effect of #1 will be most significant for the very central region where the PSF considerably smears the X-ray isophotes (cf. Figure 2) and perhaps the outermost regions where the ellipticity of the gas is poorly constrained (see below). Given the noticeable flattening of the isophotes outside the core, effect #2 will be important for $r \gtrsim 60''$. Although our analyses of the total mass and dark matter distributions in §4. and §5. do not demand $\epsilon_M = \epsilon$, we do require the value of $\epsilon_M$ computed from the data represent the same quantity when computed from the models. Unlike the models, the image contains Poisson noise. In addition to causing effect #1, the noise may also create local stable points in the image not present in the model which could yield erroneous results. In order to understand how to best treat this effect, we followed Carter & Metcalfe and generated a series of simulated images with surface density,

$$\Sigma(x, y) \propto \left( r_c^2 + x^2 + \frac{y^2}{q^2} \right)^{-1},\tag{8}$$

where $q = 1 - \epsilon$ is the ratio of the semi-minor to the semi-major axis. Poisson noise was added to these distributions with total counts comparable to the PSPC image. We held fixed the length of the semi-major axis and varied the size of the pixels in the simulated images. The results from the simulations demonstrate that the importance of effects 1 & 2, as well as the stability of the iterative procedure, depend on the size of the pixels. For images with a small number of pixels (coarse grid) the iteration becomes unstable and, if indeed it converges, converges to a value of $\epsilon_M$ usually unrelated to $\epsilon$. In contrast, when the image has too large a number of pixels (fine grid) the surface density of the image becomes very flat and the value of $\epsilon_M$ does not significantly vary from the initial guess; i.e. for the initial circular aperture, $\epsilon_M$ does not stray far from zero, regardless of the intrinsic ellipticity of the distribution. Hence one must find the pixel scale which balances the need for sufficient number of pixels to promote convergence stability while also maintaining reasonable signal-to-noise levels in each pixel.

We adopted a simple test for determination of this optimum pixel scale. For a given semi-major axis, we began by computing $\epsilon_M$ in the manner described above; i.e. start with a circular aperture and iterate until $\epsilon_M$ converges to within a desired tolerance. Then we repeated the iteration with the initial aperture shape set to a finite value of $\epsilon_M$. This yields another, possibly different, measurement of $\epsilon_M$. The spread in these values computed for many different initial $\epsilon_M$'s is a measure of the importance of the systematic errors discussed above. On performing these calculations for several semi-major axes using different pixel scales, we selected $5''$ for the pixel scale which simultaneously minimized this systematic uncertainty and the statistical error $\Delta\epsilon_M$; at this scale this systematic uncertainty is typically $< 0.02$ while for the $15''$ pixel image it is $< 0.04$. Hence we reduce systematic uncertainties associated with $\epsilon_M$ by using the image prepared as in §2.1. except that the pixels are $5''$.

In addition to the iterative moments, we also parameterize the shape of the X-ray surface brightness by fitting perfect ellipses to the isophotes following Jedrzejewski (1987; implemented with the *ellipse* task in the IRAF-STSDAS software). This method has the advantage that the computed parameters for ellipticity ($\epsilon_{iso}$) and position angle ($\theta_{iso}$) correspond to an elliptical isophote at a specific radius and thus may provide a more accurate representation of the radial variation in shape and orientation of the surface brightness. Unfortunately this technique, which was developed to study slight departures of optical isophotes from true ellipses, has the disadvantage of having larger statistical uncertainties than the iterative moments; i.e. as applied in STSDAS, the counts associated with fitting an isophote are only a small fraction of those present in the elliptical apertures used to compute the iterative moments. Because of the premium placed on counts, the image with $15''$ pixels was used for the ellipse fitting.

We list the ellipticity results in Table 3 and the corresponding position angles in Table 4 for both the iterative moments (computed for an elliptical aperture and an elliptical annular aperture) and the fitted elliptical isophotes; note that these ellipticities include the blurring due to the PSPC PSF which we will account for in our models in the later sections. The statistical uncertainties associated with the iterative moments ($\Delta\epsilon_M$) represent 90% confidence estimates while those of the fitted isophotes ($\Delta\epsilon_{iso}$) reflect 68% values; we note that the listed values of $\Delta\epsilon_M$ agree well with uncertainties estimated from the above Monte Carlo simulations. For each method the ellipticity parameters agree at all $r$ within their statistical uncertainties, where $r = (ab)^{1/2}$ is the average radius of an ellipse having semi-major axis $a$ and semi-minor axis $b$. Interior to $\sim 30''$, the X-ray isophotes are nearly circular. This could simply result from the circularly-symmetric blurring of the PSF, an effect that is reproduced in the Monte Carlo simulations when a constant ellipticity surface density is convolved with the PSF of the PSPC; this circularity implies that asymmetries due to errors of aspect correction must be quite small. The isophotes become flatter at large radii, reaching a maximum $\epsilon_M \sim 0.25$ at $r \sim 75''$. Constraints on the flattening for $r$ greater than $\sim 100''$ become weaker as the average pixel S/N approaches unity. The isophote fitting, which is most sensitive to the surface brightness S/N, does not provide meaningful ellipticity limits for $r$ greater than $\sim 90''$. Using the full elliptical aperture, $\epsilon_M \sim 0.15$ for $r \sim 110'' - 140''$ but the lower limit is only 0.06 (90% confidence). At the same large distance from the galaxy center the elliptical annulus computed on the $15''$ pixel image yields $\epsilon_M = 0.13 \pm 0.07$, which has the same statistical uncertainty associated with the full ellipse even though the annulus has $\sim 1/4$ the number of counts. Although the systematic uncertainties become larger at these distances, the Monte Carlo simulations show that systematic effects tend to (but do not always!) underestimate the true ellipticity; i.e. lower limits on $\epsilon_M$ derived using the statistical uncertainty are very likely to be conservative estimates. As a result, we measure $\epsilon_M$ for as large a distance as possible using the elliptical aperture on the $15''$ image. As a conservative estimate for the outer radius of detectable flattening, we only extend the aperture out to the distance where the "systematic variance" becomes the same magnitude as $\Delta\epsilon_M$ and the position angle agrees with the inner isophotes within uncertainties (cf. Table 4). At this distance, $r \sim 200''$ and $\epsilon_M = 0.15 - 0.25$ (90% confidence). Of course, $\epsilon_M$ computed for the whole elliptical aperture does not exactly correspond to the isophote at that distance, but comparison to the other measurements of the ellipse and annulus in Table 3 suggests that an ellipticity of at least 0.12 for $r \sim 200''$ is not unreasonable; although at these large distances the effects from the exclusion of the embedded point sources may become significant. Thus the X-ray isophotes appear to be significantly flattened out to average radius $150''$ and probably as far as $200''$.

There is no evidence for any position angle twists, although the statistical uncertainties are large for both small and large radii. The values for $\Delta\theta_M$ agree with the Monte Carlo simulations provided $\epsilon_M \gtrsim 0.15$; when the measured ellipticity is smaller, the position angle uncertainties obtained from the simulations are typically two to three times larger than the statistical estimates. We adopt the average isophote position angle $\theta_{xray} \sim 114°$ for $r \sim 70'' - 90''$ where $\Delta\theta_M$ is smallest, $\epsilon_M$ is greatest, and there is optimal agreement between all methods.

We examine the possibility that the measured ellipticities and position angles are actually caused by contamination from unresolved point sources. The centroid position of the fitted isophotes as a function of radius is a sensitive diagnostic of the presence of any substructure. We find that the centroids change by less than 1 pixel ($15''$) for $a \leq 105''$ and are consistent within their $1\sigma$ errors. In order to probe local asymmetries that affect $\epsilon_M$ but not the centroid, we examine four different cuts of the image: $x \geq 0$, $x \leq 0$, $y \geq 0$, and $y \leq 0$, where we fix the origin and define the $x$ axis to align with the major axis and the $y$ axis to be the minor axis. For each region, we create a whole image by reflecting it across the major and minor axes into the other regions. We then compute $\epsilon_M$ in an elliptical aperture following the same procedure as above. The ellipticities obtained are in excellent agreement with the values listed in Table 3 within their $1\sigma$ errors. For the ellipticities having the most well determined flattening in the original image (i.e. $75'' \leq a \leq 105''$), the values for $\epsilon_M$ differ by $< 0.02$ except for the $x \leq 0$ test at $a = 105''$; in this case $\epsilon_M = 0.30$ which exceeds by $0.05$ the mean of the other regions, but is still within the $1\sigma$ error. Thus, the consistency of all the regions requires that any contamination from unresolved point sources will have to reproduce the symmetry of all four quadrants.

Another means to examine the "lumpiness" of the X-ray image is analogous to the procedure of identifying surface brightness fluctuations of optical images (e.g, Tonry, Ajhar, & Luppino 1990). We construct a model ("bmodel" task in IRAF-STSDAS) of the X-ray surface brightness in a $240''$ square region using the results of the fitted X-ray isophotes discussed above. The model is a relatively poor fit to the central $30''$ (being too flat) but adequately represents the rest of the region. The residual image obtained from subtracting this model is featureless aside from a $\sim 2\sigma$ spike in central $30''$ due to the poor fit there. Note that usually a high order polynomial is fit the the residual image and then subtracted out. Since our image already shows no significant lumpiness after subtraction of the ellipse model, this was not necessary.

To further assess possible asymmetries, we computed one dimensional projections of the image in a $240''$ box onto the major and minor axes. We plot the result in Figure 4. The projections qualitatively exhibit the behavior of a flattened ellipsoid; i.e. the minor axis projection has the highest peak and falls off more rapidly than the major axis projection, although the two distributions are not easily distinguishable in the outer regions when poisson uncertainties are taken into account. In fact, the tails of the projections may be consistent with a slight asymmetry, but the magnitude of such an effect must be small enough to be consistent with the symmetry of $\epsilon_M$ implied in the previous analysis. The symmetry displayed by these projections allows a quantitative estimate of the strength of unresolved point sources. We make the conservative estimate of 100 being the maximum counts a source might possess without being detected anywhere on this plot. Now restricting our attention to the $75'' \leq a \leq 105''$ region that contains the isophotes critical to our analyses in the following sections, we estimate that a point source with less than 50 counts will be too weak to affect the ellipticities. We identify 28 sources within the 20 arcminute radius circle of the PSPC ribs that meet these both of these criteria. This number yields a probability of 24% that one point source lies within $60'' - 105''$. However, there is only a 2% chance that two such sources, which are required by the preceding analysis, lie within $75'' \leq a \leq 105''$. Therefore, we conclude that it is very unlikely that contamination from unresolved point sources accounts for our derived ellipticities; we will be able to determine this for certain with our planned observation of NGC 720 with the ROSAT High Resolution Imager (HRI).

The optical isophotes of NGC 720 have been studied by many authors, most recently by Nieto et al. (1992); Sparks et al. (1991); Peletier et al. (1990); Capaccioli, Piotto, & Rampazzo (1988) Jedrzejewski, Davies, & Illingworth (1987); Lauer (1985a,b); and Djorgovski (1985). All of the authors employ some variation of ellipse fitting involving Fourier analyses techniques similar to Jedrzejewski (1987), and all obtain very similar results. Within the $4''$ core radius of the galaxy (e.g., Jedrzejewski et al. 1987), the isophote ellipticity has a value $\sim 0.20$, quickly rising to $\epsilon \sim 0.40$ for semi-major axes $a \sim 15''$, then slowly increasing to a maximum ellipticity $\sim 0.45$ at $a \sim 60''$ that is maintained out to the faintest isophotes $a \sim 100''$. The position angle, in contrast to the shape, maintains a constant magnitude of $\sim 142°$. In Figure 5 we plot the X-ray isophotes depicted as perfect ellipses with ellipticity $\epsilon_M$ and position angle $\theta_M$ computed with an elliptical aperture; we also include the optical isophotes using the $R$-band data from Peletier et al. (1989).

The X-ray isophotes are everywhere rounder than the optical isophotes, but interior to $60''$ the comparison is greatly affected by the PSPC PSF. The position angles of the X-ray and optical isophotes appear discrepant by approximately $30°$ with statistical uncertainty only about $\sim 15°$. However, the uncertainties are large at small radii where the X-ray isophotes are approximately circular. We have scheduled a high resolution observation with the ROSAT High Resolution Imager to determine whether the inner X-ray isophotes actually twist and align with the optical isophotes. With the PSPC data, though, we conclude that *the major axis of the X-ray isophotes is not aligned with the major axis of the optical isophotes.*

## 2.2. Spectral Analysis

The ROSAT PSPC has moderate spectral resolution with 34 bins spanning the energy range 0.1 - 2.4 keV. With the full-scale PSPC image corrected only for embedded point sources (cf. §2.1.), we extracted the source counts from a $400''$ radius circle using the IRAF-PROS software. An annulus from $600'' - 800''$ was used for estimation of the background level which we then multiplied by a normalization factor of 1.1 to account for exposure and vignetting effects. With XSPEC, we fit the background-subtracted spectrum to a single-temperature ($1T$) optically thin plasma incorporating thermal bremsstrahlung and line emission (Raymond & Smith 1977; updated to 1992 version) with interstellar absorption. The temperature, metalicity, hydrogen column density, and emission normalization were free parameters in the fits.

Table 5 summarizes the spectral data and fit results. The Raymond-Smith $1T$ model fits the data quite well. The 90%, 95%, and 99% confidence levels for the three interesting parameters (temperature, abundances, and $N_H$), are determined by the contours of constant $\chi^2_{min} + 6.25, 8.02$, and $11.3$ respectively; these contours are plotted in Figure 6. The constraints on $N_H = (0.1 - 3.2) \times 10^{20}$ cm$^{-2}$ (95% confidence) are consistent with the galactic column density $N_H = 1.4 \times 10^{20}$ cm$^{-2}$ (Stark et al. 1992). For the abundances, He was fixed at its cosmic value while the heavy element abundances (relative abundances fixed at solar) have 99% confidence limits 8% - 60% solar. The fits place stringent constraints (95% confidence) on the temperature $T_{gas} = 0.48 - 0.69$ keV ($5 - 8 \times 10^6$ K). This contrasts with the relatively poor constraints of TFC who could only set a 90% lower limit on $T_{gas} = 0.5$ keV.

Although the single-temperature model fits the data well, it is not a unique representation of the spectrum. By fitting a two-temperature model with the heavy element abundances fixed at their solar values we obtain an equally good fit (see Table 5). The model has roughly equal contributions from the low-temperature component ($T \sim 0.45$ keV) and the high-temperature component ($T \sim 1.2$ keV), but the parameters are not precisely constrained. Hence, the PSPC spectrum cannot distinguish between a $2T$ Raymond-Smith model having solar abundances and a $1T$ model with sub-solar abundances. In fact, by simulating (with XSPEC) $2T$ spectra having 100% solar abundances and the same counts and average properties as the NGC 720 spectrum, we find that a $1T$ Raymond-Smith model fit to this simulated $2T$ spectrum will yield a good fit but with the lower temperatures ($\sim 0.5$ keV) and sub-solar abundances ($\sim 20\%$) very similar to our above $1T$ results. Determination of the precise state of the gas requires superior spectral resolution which should be achieved with ASCA and AXAF.

We investigated the presence of temperature gradients by employing the same fitting procedure as above. For examination of radial gradients, we separated the $400''$ region into an inner circle ($60''$) and an outer annulus ($120'' - 400''$). The results of the fit are listed in Table 5 with only 68% confidence estimates because of the greater uncertainty due to the smaller number of counts in each region; we do not include results for the $60'' - 120''$ region because the fewer counts associated with the region yields large uncertainties in the temperature that bracket the results of the other regions. From consideration of the 68% confidence extremes, we constrain the gradient to be $-0.26 < \left( \frac{d \ln T_{gas}}{d \ln r} \right) < 0.22$ (95% confidence), where we have taken mean values of $r$ for each of the regions. If we fix $N_H$ to its Galactic value, we obtain $-0.11 < \left( \frac{d \ln T_{gas}}{d \ln r} \right) < 0.16$ at 95% confidence and $-0.18 < \left( \frac{d \ln T_{gas}}{d \ln r} \right) < 0.21$ at 99% confidence.

In order to set more stringent limits on radial temperature gradients, we apply a Kolmogorov-Smirnov (K-S) test to the spectra of the two regions (we omitted the 0.1-0.2 keV bins because those are most subject to background uncertainties). The K-S test yields a probability of 30% that the two regions are derived from the same population. This relatively large probability serves as a discriminator for models with steep temperature gradients. To see how sensitive such a test would be in detecting a real temperature gradient we simulate Raymond-Smith spectra (with XSPEC) with statistics appropriate for the PSPC observation of NGC 720; in each region the simulated spectra have Galactic $N_H$ and 50% metallicities but different temperatures. We find that for a temperature in the inner region of $T_{in} = 0.60$ keV and an outer region temperature of $T_{out} = 0.55$ keV, the K-S probability is 15%. However, for a slightly larger gradient (i.e. $T_{in} = 0.60$ keV, $T_{out} = 0.50$ keV), the probability is reduced to 1%. We define spectral models to be inconsistent with the data if in the two regions their K-S probability is $< 1\%$; i.e. greater than a 99% discrepancy. With this criterion, we determine that for reasonable temperature ranges (i.e. $0.3 \sim 1$ keV), the temperature gradient is very precisely constrained to be $\left| \frac{d \ln T_{gas}}{d \ln r} \right| < 0.05$. Hence we find no evidence for significant radial temperature variations

Since azimuthal temperature variations might confuse the interpretation of isophote shapes

(cf. §3.1.), we also test for azimuthal gradients. We sliced the $400''$ circle into 4 equal wedges of $90°$. We defined the edges of the wedges with respect to the major axis to be (1) $-45°$ to $+45°$, (2) $+45°$ to $+135°$, (3) $+135°$ to $+225°$, and (4) $+225°$ to $-45°$; the major axis is taken along P.A. $114°$. We grouped wedges (2) and (4) into a region denoted (A) and regions (1) and (3) were grouped into region (B) in order to improve the statistics. The results of the fits for these regions are listed in Table 5. We find no evidence for a temperature gradient between (A) and (B) and set a 68% confidence upper limit $\Delta T_{gas} = T_A - T_B < 0.2$ keV. A K-S test of (A) and (B) (0.1 - 0.3 keV bins omitted) yields a probability of 70% that the two regions are derived from the same population.

## 3. Geometrical Evidence for the Existence of Dark Matter

### 3.1. Physical Interpretation of the X-ray Isophote Shapes

For the sake of clarity we summarize the physical arguments demonstrating why the X-ray emission traces the three dimensional shape of the gravitational potential. From this property we argue that the X-ray isophotes must, to a good approximation, trace the shape of the projected potential. We then show that this correspondence provides a test for dark matter independent of the gas pressure, and thus independent of the temperature profile of the gas. Finally, we discuss the validity of our assumption of quasi-hydrostatic equilibrium in NGC 720.

Since the sound crossing time for normal galaxies is much less than a Hubble time, and any bulk flows are generally less than the sound speed, the hot gas in elliptical galaxies is, to a good approximation, in a state of quasi-hydrostatic equilibrium with the underlying gravitational potential (e.g., Sarazin 1986; Binney & Tremaine 1987); i.e. $\nabla p_{gas} = -\rho_{gas} \nabla \Phi$, where $p_{gas}$ is the gas pressure, $\rho_{gas}$ is the gas mass density, and $\Phi$ is the gravitational potential. Taking the curl of this equation, one obtains $(\nabla \rho_{gas}) \times (\nabla \Phi) = 0$; surfaces of constant $\rho_{gas}$ are surfaces of constant $\Phi$, and thus the X-ray gas density "traces" the shape of the gravitational potential. One does not directly observe the gas density but instead the thermal emission from bremsstrahlung and line emission with volume emissivity (erg cm$^{-3}$ s$^{-2}$),

$$j_{gas} = n_e n_H \Lambda_{PSPC}(T_{gas}) = 0.22 \left( \frac{\rho_{gas}}{\mu m_p} \right)^2 \Lambda_{PSPC}(T_{gas}); \qquad (9)$$

where $\Lambda_{PSPC}$ is the plasma emissivity convolved with the PSPC spectral response in the hard band (0.4 - 2.4 keV), $n_e$ is the electron number density, $n_H$ is the number density of hydrogen atoms, and $\mu$ is the mean atomic weight; the coefficient 0.22 is determined assuming a completely ionized plasma with cosmic abundances. $\Lambda_{PSPC}$ is a relatively weak function of temperature. Assuming the X-rays in NGC 720 come from hot gas (see below) the range of $N_H$, abundances, and $T_{gas}$ obtained from the spectrum (§2.2.) imply that $\Lambda_{PSPC}(T_{gas})$ may only vary by $< 15\%$ (cf.

NRA 91-OSSA-3, appendix F, ROSAT mission description, Figure 10.9, 1991); i.e. $\Lambda_{PSPC}$ may be considered constant throughout the galaxy.

We may exploit quasi-hydrostatic equilibrium to relate the three dimensional shape of $j_{gas}$ to $\Phi$. Since $\rho_{gas}$ and $\Phi$ are constant on isopotential surfaces, the hydrostatic equation implies that $p_{gas}$ must also be constant on the isopotential surfaces. For an ideal gas $p_{gas} \propto \rho_{gas} T_{gas}$ then implies that surfaces of constant $T_{gas}$ are also surfaces of constant $\Phi$. Since $(\nabla \rho_{gas}) \times (\nabla \Phi) = 0$ implies that $\left(\nabla \rho_{gas}^2\right) \times (\nabla \Phi) = 0$, the quantity $\rho_{gas}^2 \Lambda_{PSPC}(T_{gas})$, and hence $j_{gas}$, must also trace the three dimensional shape of the gravitational potential.

One obtains the X-ray surface brightness ($\Sigma_X$) by projecting $j_{gas}$ onto the plane of the sky. Although $j_{gas}$ and $\Phi$ have the same three dimensional shapes, it is not true in general that $\Sigma_X$ has the same shape as the projected potential. For the case where $\Phi$ is stratified on concentric similar ellipsoids, $\Sigma_X$ and the projection of $\Phi$ have exactly the same shapes, independent of their three dimensional radial distributions (Stark 1977; Binney 1985; Franx 1988). This is not exactly true for potentials whose shape change with radius. However, one would expect that for small gradients in ellipticity, the projected shapes should closely approximate the similar ellipsoid case.

We have investigated the typical magnitude of such departures by studying simple spheroidal models whose ellipticity varies with radius. In Appendix B. we examine the projected shapes of functions having the same three dimensional shapes and radial slopes appropriate for physical potentials and gas emissivities; we include the specific example of $j_{gas}$ and the potential of the visible stars for NGC 720. We find that for reasonable ellipticity gradients, the ellipticities of the projected distributions differ by no more than $\sim 0.04$. For the above mentioned special case for NGC 720 we find that ellipticities differ by no more than $\sim 0.02$. We conclude that *to a good approximation the X-ray isophotes trace the shape of the projected gravitational potential.*

This propinquity of the shapes of the X-ray isophotes and projected potential contours enables one to assess the validity of any model for the three dimensional potential, independent of the gas pressure and temperature. In particular, we may test whether the potential due to the visible stars can produce the observed shapes of the X-ray isophotes. The only assumptions involved are the choice of the form of the deprojected potential of the visible stars and that quasi-hydrostatic equilibrium is a suitable description of the gas. We perform this test in the following section.

In order to determine whether quasi-hydrostatic equilibrium is a suitable description of the gas in NGC 720, we first mention that the X-ray emission is clearly not the result of discrete sources in the galaxy because the X-ray isophotes do not follow the shape of the optical isophotes (§2.1.2.). This property, when coupled with the good fits of the Raymond-Smith model to the spectrum (§2.2.) and the temperatures derived from them, suggests that the dominant component of the X-ray emission (0.2 - 2.4 keV) from NGC 720 is in the form of hot gas (e.g., Canizares, Fabbiano, & Trinchieri 1987). A possible complication arises since the PSPC spectrum does not rule out a multi-phase medium having cool, dense gas clouds embedded in the hot gas that are not hydrostatically supported. Thomas, Fabian, & Nulsen (1987; Thomas 1988) demonstrated that the mean density and temperature are good descriptions of a multi-phase medium, indicating that a single phase representation of the data should not significantly affect interpretation of the isophote shapes; this may not be true within the inner regions of a strong cooling flow (Tsai 1994), but the isophotes crucial to our analysis are located at relatively large distances ($\sim 10h_{80}$ kpc) which should lie safely outside the possible cooling-flow-dominated region. One must also consider possible environmental effects. In particular, the shapes of the X-ray isophotes could be distorted by either the gravitational field of a large neighboring galaxy or by ram-pressure stripping if the galaxy is traveling through a dense intergalactic medium (IGM). As indicated by Dressler, Schechter, & Rose (1986), NGC 720 has six faint companions within a 1.5 degree square field, but is quite isolated from other normal galaxies; the closest galaxy with a measured redshift lies 73 arcminutes NW. Since the presence of a dense IGM is generally associated with rich clusters of galaxies, it is unlikely that there exist significant ram-pressure distortions of NGC 720. In support of this assessment, the isophote centroids and position angles do not exhibit discernible variations with radius. We conclude that the X-ray gas in NGC 720 traces the shape of the underlying gravitational potential.

In principle this halcyon description of quasi-hydrostatic equilibrium may be corrupted by significant bulk motions of the gas; we emphasize that perfect hydrostatic equilibrium is not required, simply that additional gas motions are dynamically small. N-body simulations of hot gas in clusters of galaxies do not show evidence of large streaming motions (e.g., Tsai, Katz, & Bertschinger 1993). One would expect that any streaming motions in ellipticals would be be even less significant than in galaxy clusters since ellipticals are more relaxed systems than galaxy clusters. We are currently investigating the viability of recovering the shape of the gravitational potential by assuming quasi-hydrostatic equilibrium using N-body simulations (Buote, Tsai, & Canizares 1994, in preparation).

Although streaming may be unimportant in the gas, significant rotation of the gas also could affect its shape. That is, the gas could be flattened because it is spinning, not because the gravitational potential is flattened; i.e. another term must be included in the hydrostatic equation and thus the gas density no longer exactly traces the gravitational potential. NGC 720, like most giant ellipticals, is slowly rotating; the visible stars have a mass-weighted (optical) rotational velocity of 35 km s$^{-1}$ (Busarello, Longo, & Feoli 1992). Using the tensor virial theorem, we conclude that mass-weighted rotational velocities in excess of 150 km s$^{-1}$ are required to flatten the gas to an ellipticity of 0.25, that being the shape of the best-determined X-ray isophotes; note that in the application of the tensor virial theorum we take $W_{ii} = \int \rho_{gas} x_i \frac{\partial \Phi_{stars}}{\partial x_i} d^3x$ (no sum), where $\rho_{gas}$ is the gas density from §5.2., $\Phi_{stars}$ is the potential inferred from a constant mass-to-light ratio model in §5.1., and the integral is evaluated over the volume of the gas spheroid. Unfortunately, the PSPC, as well as all current X-ray instruments, lack the spectral resolution to detect rotation. As a result, we must resort to indirect arguments involving the properties of the visible stars and the likely history of the gas in order to assess the importance of rotation. Since the gas mass of NGC 720 (cf. §5.2.) is a small fraction of the stellar mass loss over a Hubble time as is true for most ellipticals (Mathews 1990), the gas should have net angular momentum comparable to that of the stars. Hence, if the gas rotates as fast as the stars, then the rotation is dynamically insignificant. Kley & Mathews (1993), who use hydrodynamic models of gas in elliptical galaxies to demonstrate that cooling gas eventually forms a spinning disk, emphasize that the key to forming disks lies in the fact that although the stellar rotation at any given radius may be dynamically small, conservation of angular momentum can drive up the speed of gas as it falls in to the center of the galaxy. For this scenario to be important for our analysis of NGC 720, then there must have been a significant amount of gas that has fallen in from very large radii and have been deposited at a radius $\sim 90''$. However, about 70% of the mass in visible stars is within $90''$ of NGC 720 indicating that there is insufficient stellar mass at the large radii ($r > 100''$) required to account for such rapidly rotating gas, certainly in quantities to significantly affect the observed isophotes. Therefore, the effects of rotation should not be important for the gas in NGC 720, although we can not rule it out categorically.

## 3.2. Geometric Implications

We now utilize the results of the previous section to determine whether the shapes of the X-ray isophotes are consistent with the assumption that the gas is in quasi-hydrostatic equilibrium with the visible stellar potential; note that the results of this section are intended to be primarily qualitative, we present the detailed modeling of the system in §4. and §5.. In particular, by exploiting the geometrical properties of ellipsoidal potentials, we investigate whether the stellar mass, which is much more centrally condensed than the X-ray emission, can generate the observed flattening of the X-ray isophotes (for discussions of ellipsoidal potentials see Chandrasekhar 1969; Binney & Tremaine 1987). Because the equipotential surfaces exterior to a thin homoeoid (i.e. ellipsoidal shell) are ellipsoids confocal to the homoeoid (independent of its mass), the potential becomes rounder with increasing distance from the homoeoid; interior to the homoeoid, the potential is constant. It follows that an ellipsoidal mass constructed from the sum of similar thin homoeoids will also produce a potential that becomes rounder with distance (assuming the mass density decreases with distance). When the mass is expressed as a multipole expansion, this result simply reflects the increasing importance of the monopole term with increasing distance from the center of mass.

Assuming the stellar mass is proportional to the stellar light, the results of the previous section show that we may directly compare the shapes of the projected potential surfaces produced by the stars to the observed X-ray isophotes; the effects of self-gravitation of the gas is negligible (cf. §5.2.) and we neglect it in the following discussion. We show in §5.1. that the stellar luminosity density, and hence the stellar mass density, has a radial dependence $\sim r^{-2.6}$ and core radius $r_c \sim 4''$; we take the isodensity surfaces of the stellar ellipsoid to be similar oblate spheroids having $\epsilon_{stars} = 0.40$. The stellar mass is thus considerably more centrally condensed than the X-ray gas for which $r_c \sim 16''$ and $\rho_{gas} \sim r^{-3/2}$ (cf. §5.2.); yet the X-ray isophotes display significant elongation out to $\sim 25$ optical core radii. Listed in Table 6 are the ellipticities of the stellar isopotential surfaces ($\epsilon_{pot}$) for (1) three dimensions, (2) projected onto the plane of the sky assuming the symmetry axis lies in the sky plane (cf. §4.1.), and (3) projected and convolved with the PSPC PSF; in Figure 7 we plot the projected, convolved equipotentials superimposed on the X-ray isophotes. For semi-major axis $105''$, that being the most distant isophote whose shape is very accurately determined, $\epsilon_{pot} \sim 0.10$ is much rounder than the 90% confidence lower limit of the X-ray surface brightness ($\epsilon_M \geq 0.20$); note that for $\epsilon_{stars} = 0.50$, an ellipticity greater than any of the optical isophotes, we obtain $\epsilon_{pot} \sim 0.13$ at $a = 105''$, which is still significantly less than $\epsilon_M$. If the stellar density is instead assumed to be prolate with the same radial dependence, core radius, and ellipticity as the oblate case, then the 3-D prolate potential is flatter than the oblate case at all radii by $\epsilon \sim 0.015$. The projected ellipticities of the prolate spheroid agree very well with the results for the oblate case as is expected since the distinction between prolate and oblate spheroids having ellipticities $\sim 0.10$ is not large. We also list in Table 6 the ellipticity of the X-ray isophotes ($\epsilon_{isophote}$) predicted from our detailed models of gas in the stellar potential (§5.) assuming the gas is isothermal and ideal; note the excellent correspondence between the projected potential ellipticity and the isothermal isophotes. The discrepancy between the expected shape of the stellar potential and the observed X-ray isophotes is actually amplified because a roughly uniform background will tend to decrease the measured values of $\epsilon_M$ for the X-ray isophotes (Carter & Metcalfe 1980).

We quantify the reality of this inconsistency with Monte Carlo simulations using the pseudo spheroids discussed in Appendix B.. In the notation of Appendix B., we assume the gas emissivity $j_{gas} \propto (a_0^2 + \xi^2)^{-3/2}$ with the same $\epsilon(r)$ as $\Phi_{stars}$. We project $j_{gas}$ onto the plane of the sky and convolve with the PSPC PSF. Then Poisson counts appropriate to the NGC 720 PSPC observation and a uniform background are added to simulate an observation. After subtracting out a uniform background, the ellipticities are then computed using the iterative moment technique with a circular aperture as described in §2.1.2.. In Table 7 we list the results of 1000 simulations for both the oblate and prolate constant mass-to-light ratio models. As expected, the lower bounds on $\epsilon_M$ are near 0 as a result of the systematic effects discussed in §2.1.2.. However, the upper bounds also show large departures from the mean. For semi-major axis $105''$, the value of $\epsilon_M$ in the simulations is as large as that measured from the real X-ray data isophotes ($\epsilon_M = 0.25$) in only 1% of the simulations. These simple models demonstrate that the constant mass-to-light ratio models are inconsistent with the observed flattening of the X-ray isophotes at the 99% confidence level; even upon considering the maximum uncertainty due to comparing the projections of non-similar spheroids (cf. Appendix B.), the discrepancy is still robust at the 90% level. This discrepancy is also unlikely due to possible rotation of the gas since upon adding a uniform rotation term to the constant $M/L$ potential we find that mass-weighted velocities $v_0 > 120$ km s$^{-1}$ are required to produce the X-ray ellipticities; such velocities are significantly larger than expected from the stellar rotation and are consistent with the velocities required from the tensor virial theorum obtained in the previous section. Hence, by employing simple arguments involving the properties of spheroids and their potentials, we conclude that an spheroidal mass distribution confined to the shape of the stellar matter cannot produce a gravitational potential flat enough to yield the observed ellipticities of the X-ray isophotes; a conclusion that is independent of the pressure and temperature of the gas, and the amount of stellar mass. Assuming the gas is in quasi-hydrostatic equilibrium (cf. previous section), then *there must exist in NGC 720 an extended halo of dark matter sufficiently elongated to account for the isophotal flattening.*

In the previous discussion we have ignored the position angle offset of the optical and X-ray major axes. If the stars are solely responsible for the gravitational potential, the gas and stars must be co-axial. In addition, if the stellar ellipsoid is axisymmetric, so must be the gas with the same type of axisymmetry; i.e. if the stars are oblate, the potential, and hence the gas, must also be oblate. For this case there can be no apparent position angle misalignments due to projection on the sky, regardless of any intrinsic variations of ellipticity with radius (e.g., Mihalas & Binney 1981). If the stars are indeed triaxial, then a projected misalignment of the X-ray and optical major axes is theoretically possible. Detailed triaxial models would be required to see if triaxiality can actually explain the observed offset without dark matter. This is moot, given our conclusion that the shape itself requires dark matter, but we will examine such models in a future paper that will include a ROSAT High Resolution Imager (HRI) observation of NGC 720 (see Bertola et. al. 1991 for an example of this problem).

It is difficult to quantify exactly the expected uncertainty associated with the position angle from the above Monte Carlo simulations because when $\epsilon_M \sim 0$, the position angle is not well defined. However, when selecting only those runs where, say, $\epsilon_M \geq 0.10$ ($\sim 400 - 500$ simulations), the position angle uncertainty is $\sim 9°$ at 68% confidence, $\sim 15°$ at 90% confidence, $\sim 20°$ at 95% confidence, and $\sim 27°$ at 99% confidence. For more elongated $\epsilon_M$, the uncertainty is even smaller. The position angle discrepancy between the stellar and X-ray isophotes adopted in §2.1.2. is 28°. For the elongated X-ray isophotes, the position angle implied by the simple models for the visible stellar mass is inconsistent with the observed values at the $\sim 99\%$ level. Hence the offset of the major axes may provide further geometrical evidence for the existence of unseen matter in NGC 720.

### 3.3. Implications for Alternative Theories of Gravitation

The geometrical test for dark matter introduced in the previous section places new constraints on theories of generalized forces. Instead of invoking the existence of unseen mass to explain the flat rotation curves in spiral galaxies, these theories modify the Newtonian force law in such a manner to account for the observed gravitational effects; see Liboff (1992) for a concise summary of this subject and Sanders (1990) for a more extensive review. Perhaps the most successful of these theories is the "Modification of Newtonian Dynamics" (MOND) proposed by Milgrom (1984a,b,c,1986). Milgrom proffers that the gravitational acceleration ($\vec{g}_M$) due to a point mass,

$M$, is characterized by,

$$\vec{g}_M = -\frac{GM}{r^2}\hat{r}, \quad \text{for} \quad |\vec{g}_M| \gg a_0, \tag{10}$$

and

$$\vec{g}_M = -\frac{\sqrt{GMa_0}}{r}\hat{r}, \quad \text{for} \quad |\vec{g}_M| \ll a_0, \tag{11}$$

where $a_0$ is the appropriate acceleration scale that yields circular velocities ($v_c \propto M^{1/4}$) consistent with observations of the infrared Tully-Fisher relation for spiral galaxies if $M \propto L$. Bekenstein & Milgrom (1984) formulate MOND as a nonrelativistic potential theory for gravity for which they obtain a field equation,

$$\nabla \cdot [\mu(x)\nabla\Phi_M] = 4\pi G\rho, \tag{12}$$

where $\Phi_M$ is given by $g_M = -\nabla\Phi_M$, $x = |\nabla\Phi_M|/a_0$, and $\mu(x)$ is some unspecified smooth function (assumed monotonic) appropriately connecting the Newtonian and Milgrom domains; note that this equation is non-linear and thus the principle of linear superposition is not obeyed by MOND.

By exploiting the region in the galaxy where Newtonian gravity applies to high precision (i.e. $g/a_0 \gg 1$), we may obtain robust constraints on the shape of the MOND potential produced by the stars without solving the non-linear field equation (eq. [12]). Consider the MOND potential expressed in terms of spherical harmonics,

$$\Phi(r,\theta,\phi) = \sum_{l,m,i} A^i_{lm}(r)Y^i_{lm}(\theta,\phi), \tag{13}$$

where $Y^i_{lm}$ is the spherical harmonic of order $l, m$ with $i$ indicating whether it is even or odd in $\phi$. For an arbitrary mass distribution, Milgrom (1986) demonstrates that for $l \neq 0$, $A^i_{lm}(r) \to a^i_{lm}r^{-\xi_l}$ in the limit $r \to \infty$, where $\xi_l = [l(l+1)/2]^{1/2}$ and the $a^i_{lm}$ are constants; the $l = 0$ "monopole" term is the spherically-symmetric $\Phi_0(r) = \sqrt{GMa_0}\ln(r)$. It follows then that the $l$th multipole of MOND decays slower than in the Newtonian theory ($r^{-(l+1)}$), but the spherically-symmetric monopole term does indeed eventually dominate at large distances; i.e. the MOND potential becomes more spherical with distance just as in Newtonian theory (provided, of course, the density is monotonically decreasing). As a result, we have a qualitative description linking the Newtonian and Milgrom regimes: the ellipticity of the potential generated by the stars in the region where Newtonian physics applies serves as an upper limit to the ellipticity at larger distances because the potential must become rounder with increasing distance, albeit more slowly in the MOND regime.

In order to set a realistic upper bound on the potential shapes, we need to properly define the "Newtonian Regime". Milgrom (1986) defines the transition radius $r_t \equiv (GM/a_0)^{1/2}$ between the Newton and Milgrom regions where $M$ is taken to be the total mass of the bound system. In his review, Sanders (1990; Begeman, Broeils, & Sanders 1991) shows that $a_0 \approx 10^{-8}$ cm s$^{-2}$ ($H_0 = 75$ km s$^{-1}$ Mpc$^{-1}$) in order to explain the flat rotation curves of spiral galaxies. Assuming that the stars constitute the only mass in NGC 720, then $r_t \sim 12$ kpc, where we have used $10^{11} M_\odot$ (cf. §5.1.) for the stellar mass and a distance of 21 Mpc ($H_0 = 75$ km s$^{-1}$ Mpc$^{-1}$). This transition distance is consistent with previous estimates which place $r_t$ between 10 and 20 kpc (cf. Sanders 1990; Liboff 1992). Expressing $r_t$ in arcseconds, we have in the context of MOND that the Newtonian regime applies for $r < r_t = 120''$.

The analysis of the previous section (§3.2.) may be carried over in totality because the relevant X-ray isophotes have semi-major axes $a \sim 100''$; i.e. our geometrical discussion lay entirely in the Newtonian regime. In fact, our analysis applies even when restricted to a smaller region where presumably the Newtonian approximation is even a better description. If we use $a = 30''$ (3 kpc), for example, as a reference, the projected ellipticity of the stellar potential is 0.13 (cf. Table 6). This value is already rounder than the X-ray isophotes at $a = 105''$ and the discrepancy must be amplified for the stellar isopotential at $a = 105''$ since the ellipticity of the MOND potential must decrease with distance. We conclude that *MOND does not obviate the need for dark matter because the stellar potential is already too round to explain the observed flattening of the X-ray isophotes in the region where Newtonian physics would still apply.*

We may also examine MOND without reference to the actual value of $a_0$. Equation (12) may be expressed in terms of the Newtonian field $\vec{g}_N = -\nabla \Phi_N$,

$$\mu(g_M/a_0)\vec{g}_M = \vec{g}_N + \nabla \times \vec{h} \tag{14}$$

where $\vec{g}_M = -\nabla \Phi_M$ is the MOND gravitational field, and $\vec{h}$ is an unspecified field (Bekenstein & Milgrom 1984). In order to satisfy the basic assumptions of MOND expressed by equations (10) and (11), the curl term in equation (14) must be small with respect to $\vec{g}_N$; Bekenstein & Milgrom do show that $\nabla \times \vec{h}$ decreases faster than $\vec{g}_N$ at large distances. For systems possessing a high degree of symmetry (e.g., spherical, planer, and cylindrical), the curl term vanishes exactly. Hence, $\mu(g_M/a_0)\vec{g}_M \approx \vec{g}_N$ must be a good approximation for an arbitrary system if indeed the field equation is to reproduce the basic tenets of MOND and connect appropriately to Newtonian mechanics.

Equation (14) implies that for a surface where $g_N = constant$, $g_M$ must also be nearly constant, and thus $\vec{g}_M/\vec{g}_N$ is also constant; i.e. surfaces of constant acceleration in MOND are approximately surfaces of constant acceleration in Newtonian gravitation. Applying this approximation of MOND to the stellar matter distribution of NGC 720 yields the same isopotential shapes derived for the Newtonian case discussed in §3.2.; we are currently examining numerical solutions of the field equation to obtain shape constraints on the MOND potential to arbitrary accuracy (Bertschinger, Buote, & Canizares 1993, in preparation). *To the accuracy implied by taking $\mu(g/a_0)\vec{g} \approx \vec{g}_N$, MOND cannot account for the observed flattening of the X-ray isophotes without invoking the existence of dark matter, independent of the value of $a_0$*

## 4. Total Gravitating Matter Distribution

### 4.1. Model

We investigate how the the the morphology of the X-ray gas constrains the structure of the total galaxian mass. Except for some minor improvements, we employ the technique described by Buote & Canizares (1992; Buote 1992) which involves four principal steps: (1) modeling the gravitational potential, (2) "filling" the potential well with hot, X-ray emitting gas, (3) projecting the emission onto the plane of the sky, and (4) convolving the emission with the PSPC PSF to compare to observations.

We assume the gross structure of the mass is adequately described by a single ellipsoid of constant shape and orientation; in §5. the contributions from the stars, X-ray gas, and dark matter will be analyzed separately; in a future paper we will explore the effects of other types of mass models. We consider mass densities of both Ferrers (cf. Chandrasekhar 1969) and Hernquist (1990) types. For an ellipsoid having semi-axes $a_i$, the Ferrers (i.e. power-law) density has the dimensionless form,

$$\tilde{\rho}_F(\vec{x}) = \left[ \left( \frac{a_0}{a_3} \right)^2 + m^2 \right]^{-n}, \qquad m^2 = \sum_{i=1}^{3} \frac{x_i^2}{a_i^2}, \tag{15}$$

where $a_0$ is the core parameter, $a_3$ is the semi-major axis, and the dimensionless number $m$ defines the equation of a homoeoid between the origin ($m = 0$) and the boundary ($m = 1$) of the ellipsoid. As discussed in Binney & Tremaine (1987), power-law densities having $2 < 2n < 3$ are suitable approximations of the mass and light profiles of many galaxies. Applying the notation of equation (15), the dimensionless Hernquist (1990) density becomes,

$$\tilde{\rho}_H(\vec{x}) = m^{-1} \left[ \frac{a_0}{a_3} + m \right]^{-3}, \tag{16}$$

where the ellipsoidal surface enclosing half of the mass is defined by $m_{1/2} = (1 + \sqrt{2})a_0/a_3$ (Hernquist 1992) for a mass distribution extending throughout all space; equation (16) gives rise to an excellent approximation of the de Vaucouleurs $R^{1/4}$ law. In order to limit the number of free parameters in our model, we consider axisymmetric ellipsoids. The oblate spheroid has $a_1 = a_3$ and $a_2 = (1 - \epsilon)a_3$, where $\epsilon$ is the ellipticity of the isodensity surfaces in the $(x_1, x_2)$ and $(x_2, x_3)$ planes. For the prolate case, $a_1 = a_2 = (1 - \epsilon)a_3$, where $\epsilon$ is now the ellipticity in the $(x_1, x_3)$ and $(x_2, x_3)$ planes. By generating both oblate and prolate models we bracket the triaxial case (Binney & Strimple 1978).

The gravitational potential generated by these densities is a complicated function requiring numerical evaluation; for a discussion of ellipsoidal potentials see Chandrasekhar (1969) and Binney & Tremaine (1987). The potential of an ellipsoidal mass with a finite outer boundary may be written as,

$$\Phi_\alpha(\vec{x}) = -\frac{GM}{S_\alpha} \phi_\alpha(\vec{x}), \tag{17}$$

where $\alpha = F$ refers to a Ferrers density and $\alpha = H$ refers to the Hernquist density, $G$ is Newton's constant, and $M$ is the total ellipsoidal mass; $S_\alpha$ is a dimensionless number related to the mass,

$$S_\alpha = 4 \int_0^1 \tilde{\rho}_\alpha(m^2) m^2 \, dm, \tag{18}$$

where $\hat{\rho}_\alpha$ refers to either equation (15) or (16). The function $\phi(\vec{x})$ has the dimensions of inverse length and for the Ferrers density has the form,

$$\phi_F(\vec{x}) = \int_\lambda^\infty \frac{du}{\Delta} \begin{cases} \log\left[\frac{(a_0/a_3)^2+1}{(a_0/a_3)^2+m^2(u)}\right] & n=1 \\ \frac{1}{1-n}\left[\left(\left(\frac{a_0}{a_3}\right)^2+1\right)^{1-n} - \left(\left(\frac{a_0}{a_3}\right)^2+m^2(u)\right)^{1-n}\right] & n\neq 1 \end{cases}, \tag{19}$$

where

$$\Delta^2 = \prod_{i=1}^3 (a_i^2+u), \qquad m^2(u) = \sum_{i=1}^3 \frac{x_i^2}{a_i^2+u}, \tag{20}$$

and $\lambda$ is the ellipsoidal coordinate of the point $\vec{x}=(x_1,x_2,x_3)$; $\lambda$ is defined so that $m^2(\lambda)=1$ for $\vec{x}$ exterior to the bounding ellipsoid, and $\lambda=0$ for $\vec{x}$ interior to the bounding ellipsoid. The expression for $\phi(\vec{x})$ using the Hernquist density is,

$$\phi_H(\vec{x}) = \int_\lambda^\infty \frac{du}{\Delta}\left[\left(\frac{a_0}{a_3}+m(u)\right)^{-2} - \left(\frac{a_0}{a_3}+1\right)^{-2}\right]. \tag{21}$$

By normalizing $\Phi$ (eq. [17]) to its central value, we generate potential families of varying scale ($a_0$) and shape ($\epsilon$).

The potential is then "filled" with hot, X-ray emitting gas by making the fundamental assumption that the gas is in hydrostatic equilibrium with the underlying gravitational potential $\Phi$ (cf. §3.1.). If the gas is isothermal and obeys an ideal gas equation of state, then the equation of hydrostatic equilibrium may be solved exactly to give,

$$\rho_{gas}(\vec{x}) = e^{[1-\Phi(\vec{x})]\Gamma}, \qquad \Gamma = \frac{\mu m_p \Phi_0}{k_B T_{gas}}, \tag{22}$$

where $\rho_{gas}$ and $\Phi$ are normalized to their central values $\rho_{gas}(0)$ and $\Phi_0$, $\mu$ is the mean atomic weight, $m_p$ is the proton mass, $k_B$ is Boltzmann's constant, and $T_{gas}$ is the gas temperature. For a given potential shape, $\Gamma$ is well constrained by the radial profile of the X-ray surface brightness; i.e. we do not require knowledge of either the gas temperature ($T_{gas}$) or the depth of the potential ($\Phi_0$), and therefore the mass of the galaxy. In fact, results concerning the shape of the potential are not particularly sensitive to the assumption of isothermality because the PSPC is relatively insensitive to the range of $T_{gas}$ implied by the the galaxy spectrum (cf. §2.2. and §3.1.).

We test the effects of possible temperature gradients on the shape measurements. First, we consider a linear perturbation to the isothermal case,

$$T(a) = T_0\left(1 + \frac{a}{a_s}\delta\right), \tag{23}$$

where $a = m a_3$ is the elliptical radius, $a_s$ is an appropriate scale length and $\delta$ is a free parameter. For $\delta$ sufficiently small, the equation for $\rho_{gas}$ is the same as (22) except that $T_{gas}$ in $\Gamma$ is replaced with (23). Second, we consider a polytropic relation, $p_{gas} = K\rho_{gas}^\gamma$ ($K = constant$), which yields upon substitution into the hydrostatic equilibrium equation,

$$\rho_{gas} = \left[\frac{\gamma - 1}{\gamma}(\Phi - 1)\Gamma + 1\right]^{\frac{1}{\gamma - 1}};$$ (24)

where $\rho_{gas}$ and $\Phi$ are normalized to their central values, and $\Gamma = |\Phi_0|/K\rho_{gas}^{\gamma-1}(0)$. If in addition the gas is assumed to be ideal, then $\Gamma = \mu m_p |\Phi_0|/k_B T_{gas}(0)$, and the temperature is simply proportional to the expression within the brackets of equation (24).

We have shown in §3.1. that the X-ray emission of the gas is accurately represented by $\rho_{gas}^2 \times$(weak function of temperature). Hence, the surface brightness of the gas may now be constructed by simply integrating $\rho_{gas}^2$ along the line of sight,

$$\Sigma_X(y, z) \propto \int \rho_{gas}^2 dx,$$ (25)

where the $y - z$ plane coincides with the sky. This scheme assumes that the symmetry axis of the spheroid lies in the plane of the sky. Given the observed flattening of the stellar distribution of NGC 720, we believe that a substantial inclination of the symmetry axis is unlikely because (1) the observed number of galaxies flatter than NGC 720 is relatively small (Fasano & Vio 1991; Lambas, Maddox, & Loveday 1992; Ryden 1992, 1991), (2) galaxies substantially flatter than NGC 720, and not rotationally supported, are dynamically unstable (Merrit & Stiavelli 1990; Merrit & Hernquist 1991), and (3) dynamical studies of NGC 720 by Binney, Davies, & Illingworth (1990) and van der Marel (1991) suggest that the galaxy is nearly edge-on. Furthermore, we are not sensitive to small inclination angles (cf. Binney & Strimple 1978; Fabricant, Rybicki, & Gorenstein 1984; and Buote & Canizares 1992). The final step consists of convolving $\Sigma_X(y, z)$ with the PSPC PSF described in §2.1., and comparing the result to the PSPC image.

## 4.2. Shape of the Total Matter

Our procedure to determine the shape of the total matter begins by specifying the semi-major axis length ($a_3$) of the spheroid. Then, for a given total matter ellipticity ($\epsilon_{tot}$) we generate surface brightness maps for any values of $a_0$ and $\Gamma$; here we have assumed the isothermal gas solution (eq. [22]). Using a $\chi^2$ fit to compare the radial profile of the model image to the data, we obtain the 90% confidence interval ($a_0, \Gamma$) defined by those models having $\chi^2 \le \chi_{min}^2 + 4.61$; note that the models with temperature gradients have three interesting parameters (e.g., $a_0, \Gamma, \delta$) and the corresponding $\Delta\chi^2 = 6.25$ to determine the 90% confidence level. Within this 90% interval, we compute the minimum and maximum ellipticities of the model surface brightness ($\epsilon_{model}^{min}, \epsilon_{model}^{max}$) using the iterative moment technique as described in §2.1.2. for an elliptical aperture having semi-major axis 90″. The upper limit for $\epsilon_{tot}$ is obtained by finding the smallest value of $\epsilon_{tot}$ such that $\epsilon_{model}^{min} > \epsilon_{data}^{max}$ in its 90% confidence interval, where $\epsilon_{data}^{max}$ is the 90% confidence upper limit on $\epsilon_M$ from Table 3. In the same manner, a lower limit is obtained by finding the largest value of $\epsilon_{tot}$ such that $\epsilon_{model}^{max} < \epsilon_{data}^{min}$. As we discuss in §6., $\epsilon_{tot}$ is in effect constrained only out to distances where $\epsilon_{data}$ is well determined.

We list in Table 8 the results for the isothermal gas solution (eq. [22]) assuming $a_3 = 450''$; the fit results of a typical model are shown in Figure 8. The $\rho \propto r^{-2}$ and Hernquist density distributions yield excellent fits to the X-ray surface brightness while $\rho \propto r^{-3}$ is too steep to adequately reproduce the data. Each of the density profiles yields very large ellipticities for the gravitating matter with lower limits only marginally consistent with the maximum stellar isophote ellipticity of $\sim 0.45$. For smaller $a_3$, the quality of the fits diminishes for each density model, which sets a lower limit on $a_3$. We define the fits to be unacceptable if the probability that $\chi^2$ should exceed the measured value of $\chi_{min}^2$ by chance is less than 10%. In this manner we obtain lower limits on $a_3$ of $225''$ and $260''$ for the oblate and prolate ($\rho \sim r^{-2}$) cases respectively; there is no upper bound. For $a_{3min} < a_3 < 450''$, the $\epsilon_{tot}$ limits change by less than 0.01.

The results for the linearly perturbed isothermal models (eq. [23]) agree very well with the isothermal results. The best-fit values for the $\delta$ parameter are negative and have typical magnitudes $\sim 0.06$; the scale length $a_s$ is set to $400''$ in all the fits. From consideration of only the fits to the radial profile, we obtain 90% confidence limits (oblate models) of $\delta = -0.06^{+0.13}_{-0.14}$ and $\epsilon_{tot} = 0.51 - 0.79$; i.e. these models have larger parameter spaces than the isothermal models and bracket the isothermal results. However, by considering the temperature gradients implied by the expanded parameter space we may eliminate those models inconsistent with the PSPC spectrum (§2.2.). That is, we compute emission-weighted temperatures of the models in the $0'' - 60''$ and $120'' - 400''$ regions and then simulate PSPC Raymond-Smith spectra as described in §2.2.. These simulated spectra are then compared to the allowed gradients implied by the K-S results for the actual data in §2.2.. When restricting the parameter spaces to be consistent with the K-S tests, we obtain results almost identical to the isothermal case. Although our models do not account for the reduction in central temperature due to a possible cooling flow, the comparison should not be greatly affected since we average over a large region. These same results apply to the $r^{-3}$ and Hernquist models.

The polytropic equation equation of state (eq. [24]) yields results that are essentially identical to the linearly perturbed isothermal models. For the $\rho \sim r^{-2}$ model, the polytropic indices derived from the fits span the range $\gamma = 1.06^{+0.17}_{-0.20}$ for oblate models and $\gamma = 1.10^{+0.18}_{-0.19}$ for prolate models (90% confidence); the ellipticities also have a larger range than the isothermal case: $\epsilon_{tot} = 0.50 - 0.77$ for oblate models and $\epsilon_{tot} = 0.46 - 0.69$ for prolate models. However, just as with the linear temperature model, the constraints from K-S tests eliminate those models which differ significantly from the isothermal case. As a result, the polytropic models agree very well with the isothermal solution. Again, these same results apply to the $r^{-3}$ and Hernquist models.

### 4.3.  Estimate of the Total Matter

Equation (17) and the definition of $\Gamma$ (eq. [22]) combine to give an expression for the total mass,

$$M_{tot} = \frac{S_\alpha}{G\phi_\alpha(0)} \left( \frac{k_B T_{gas}}{\mu m_p} \right) \Gamma, \qquad (26)$$

where as before $\alpha$ refers to either a Ferrers or Hernquist mass profile, and $\phi$ is evaluated at the center of the spheroid. Using the 90% confidence results from Table 8 with the above equation, we list in Table 9 the total masses ($M_{tot}$) and the corresponding values of $\Upsilon_B = M_{tot}/L_B$ in solar units for both the $\rho \sim r^{-2}$ and Hernquist densities; the $B$-band luminosity $L_B = 2.2 \times 10^{10} h_{80}^{-1} L_\odot$ is obtained by scaling $B_T = 11.15$ from Burstein et al. (1987) to $D = 20 h_{80}$ Mpc; also listed are $\rho \sim r^{-2}$ results assuming the minimum acceptable semi-major axis length for the total matter spheroid. There is no significant difference in $M_{tot}$ for the $\rho \sim r^{-2}$ and Hernquist densities of the same $a_3$. However, $M_{tot}$ is systematically less for smaller $a_3$ because the density profile is essentially the same for all the cases but the total spheroidal volume is not. In Figure 9 we plot the integrated mass ($\rho_{tot} \sim r^{-2}$) interior to a spheroid of semi-major axis $a < a_3$, where $a$ is the elliptical radius defined by $a = m a_3$ and $a_3$ is the spheroid having mass $M_{tot}$. As expected, the masses for $a_3 = 450''$ and $a_3 = 225''$ demonstrate good agreement at $a = 225''$ although the $a_3 = 450''$ has systematically more mass for small $a$.

Assuming a stellar $\Upsilon_B \sim 7 \Upsilon_\odot$ (§5.1.), and neglecting the mass of the gas (i.e. $M_{tot} = M_{stars} + M_{DM}$, cf. §5.2.), we obtain 90% confidence limits on the ratio of dark matter to stellar matter, for both oblate and prolate $\rho_{tot} \sim r^{-2}$ models, of $M_{DM}/M_{stars} = 4 - 9$ at $a_3 = 450''$ and $M_{DM}/M_{stars} = 3 - 5$ at the minimum $a_3$; note that these values may be systematically low due to the uncertainty in $\Upsilon_B$ for the stars described in §5.1.. We are unable to set an upper bound on the mass because $a_3$ is not constrained by the data, but we obtain a 90% confidence lower bound $\Upsilon_B > 20 h_{80}^{-1} \Upsilon_\odot$ using the prolate $a_3 = 260''$ models which have the minimum acceptable value of $a_3$ (see above).

TFC estimate the binding mass of NGC 720 by inferring the X-ray gas density from deprojecting the spherical King function (eq. [3]) and then employing the equation of hydrostatic equilibrium (eq. [1]). Assuming the gas is isothermal with temperatures consistent with our single-temperature models in §2.2., TFC find $M_{tot} \sim 6 \times 10^{11} h_{80} M_\odot$ at $r = 240''$, in excellent agreement with our values at that distance. Binney, Davies, & Illingworth (1990; also van der Marel 1992) utilize $R$-band surface photometry and extensive spectroscopic data to generate axisymmetric mass models for NGC 720. Within $\sim 60''$, Binney et. al. obtain $\Upsilon_B < 17.2$ scaled to $D = 20 h_{80}$ Mpc, which is consistent with the values in Figure 9. They also determine that a spatially constant value of $\Upsilon_B$ is consistent with their models; we will address this issue in the following section.

Franx (1993) shows that simple models of elliptical galaxies with massive halos satisfy a Tully-Fisher relation provided $v_c/\sigma_0 \sim 1.38$, where $v_c$ is the maximum circular velocity of the halo and $\sigma_0$ is the observed central velocity dispersion. We may compute this quantity directly from our models. In cylindrical coordinates, the circular velocity for an oblate spheroid is,

$$v_c^2(R) = R \frac{\partial \Phi(R, z)}{\partial R}\bigg|_{z=0} \tag{27}$$

$$= \frac{4GM}{S_\alpha} \int_0^{R/a_3} \frac{\tilde{\rho}_\alpha(m^2) m^2 dm}{\sqrt{R^2 - e^2 a_3^2 m^2}}, \tag{28}$$

where $S_\alpha$ and $\tilde{\rho}_\alpha$ are defined by equation (18), $e = \sqrt{1 - q^2}$ is the eccentricity, and in the notation of §4.1., $R^2 = x_1^2 + x_3^2$ and $z = x_2$. Hence, we may compute $v_c/\sigma_0$ using the total masses derived above; in principle, $v_c/\sigma_0 \sim 1.38$ may be used to constrain the mass as well. By using the previously derived total masses (90% confidence) for $\rho \sim r^{-2}$ and $a_3 = 450''$, we obtain $v_c(a_3) = (327 - 409)$ km s$^{-1}$ and $v_c/\sigma_0 = 1.26 - 1.57$, where $\sigma_0 = 260$ km s$^{-1}$ (Binney et al. 1990). For the models having $a_3 = 225''$, we obtain $v_c(a_3) = (351 - 429)$ km s$^{-1}$ and $v_c/\sigma_0 = 1.35 - 1.65$. Therefore our potentials derived from analysis of the X-ray gas yield $v_c$ and $v_c/\sigma_0$ consistent with the models of Franx (1993), the agreement being better for the models having larger $a_3$.

## 5. Dark Matter Distribution

We utilize knowledge of the observed stellar and X-ray gas distributions to determine the distribution of dark matter. The total gravitational potential of the galaxy is simply,

$$\Phi = \Phi_{stars} + \Phi_{gas} + \Phi_{DM}, \tag{29}$$

where $\Phi_{stars}$, $\Phi_{gas}$, and $\Phi_{DM}$ are respectively the potentials of the visible stellar distribution, the X-ray emitting gas, and the dark matter. We would like to emphasize that $\Phi_{stars}$ is simply the potential inferred directly from the optical light (i.e. constant mass-to-light ratio model having mass of visible stars), $\Phi_{gas}$ is inferred directly from the observed X-rays, and $\Phi_{DM}$ is anything else – we do not assume anything about the composition of the dark matter, only that it is distributed differently from the visible stars and X-rays. In the notation of equation (17) we can express the above potential as,

$$\Phi = -GM_{stars} \left[ \frac{\phi_{stars}}{S_{stars}} + \frac{\phi_{gas}}{S_{gas}} \left( \frac{M_{gas}}{M_{stars}} \right) + \frac{\phi_{DM}}{S_{DM}} \left( \frac{M_{DM}}{M_{stars}} \right) \right]. \tag{30}$$

Since $\Phi$ normalized to its central value is all that is necessary to constrain its shape, the masses enter only in terms of ratios to $M_{stars}$. As we show below, the ratio $M_{gas}/M_{stars}$ is small for reasonable values of $M_{stars}$. Hence, the only free parameters of importance are those associated with the shape of the dark matter and the ratio $M_{DM}/M_{stars}$.

We emphasize that determination of $M_{DM}/M_{stars}$ by fits to the X-ray radial profile is independent of the distance to the galaxy. By comparing the total mass obtained from this method to the mass derived from the distance-dependent equation (26), one can in principle constrain the distance to the galaxy; of course, this method will depend to some extent on the functional forms assumed for the three mass components. Unfortunately the PSPC constraints on $T(r)$ are still not precise enough to set strong constraints on the mass. We must await instruments with superior spatial and spectral resolution (e.g., AXAF) to determine the viability of this method as a distance indicator.

### 5.1. Stellar Mass

We estimate the stellar mass density ($\rho_{stars}$) by assuming that it is proportional to the stellar light. Comprehensive major-axis $R$-band surface photometry data exists in the literature for NGC 720 (see references in §2.1.2.) allowing us to examine data spanning the whole galaxy; i.e. Lauer (1985) concentrates on the inner $5''$ of the galaxy; Jedrzejewski, Davies, & Illingworth (1987), who like other authors, publish data out to $\sim 60''$; and Peletier et al. (1990) who publish data for NGC 720 extending out to $\sim 120''$; see Peletier et. al. (1990) for a discussion regarding the consistency of these data sets.

For simplicity, we fit functions to the surface brightness data that are projections of either the Ferrers or Hernquist models (§4.1.). Since our models require that the mass be bounded (i.e. $0 \leq m \leq 1$), we have to arbitrarily assign an edge to the stellar matter; we also assume the galaxy is not inclined along the line of sight (cf. §4.1.). The projection of the Ferrers model proceeds by considering an oblate spheroid having semi-major axes $a_x$, $a_y$ and semi-minor axis $a_z = qa_x$, where $q$ is the axial ratio, and the $(y, z)$ plane is the sky plane. The luminosity density for the Ferrers model is then $j_F \propto (a^2 + x^2 + y^2 + z^2/q^2)^{-n}$. Since we fit only the surface brightness data on the projected major axis ($z = 0$), $j_F$ only depends on $r = \sqrt{x^2 + y^2}$. By exploiting the circular symmetry in the plane in the same manner that is done for spherical systems (cf. Binney & Tremaine 1987, §2.1 (d)), we obtain the projected luminosity,

$$I_F(R) \propto \left[1 + \left(\frac{R}{a}\right)^2\right]^{n-1/2} \int_0^B \frac{du}{(1 + u^2)^n}, \qquad B = \left(\frac{a_3^2 - R^2}{R^2 + a^2}\right)^{1/2}, \qquad (31)$$

where $R$ is the projected radius, $a_3 = a_x = a_y$ is the edge of the stellar matter, $a$, and $n$ are free parameters; note that $q$ is not constrained by this method. By fitting $I_F$ to the major axis surface brightness we obtain $j_F$, from which follows $\rho_{stars} \propto j_F$. For the projected Hernquist density, we just use the de Vaucouleurs $R^{1/4}$ Law. We assume 2% uncertainties for all the data sets

Neither of the models fit the surface brightness with high accuracy over the whole galaxy. Generally $I_F$ is an excellent description of both the inner $\sim 60''$ where $n \sim 1.25$, and outside $60''$ where $n \sim 1.5$. Fitting the whole galaxy, in contrast, yields a marginal result that is a good representation of the core, but slightly too flat in the outskirts. The $R^{1/4}$ Law characterizes well the outer regions ($\gtrsim 60''$) of the galaxy, but is a terrible fit in the interior. We choose to employ a single power law over the entire galaxy because (1) most of the light is concentrated in the regions where $I_F$ is a very accurate description of the surface brightness, and (2) the increased accuracy of a more sophisticated model (e.g., a smooth joining of the Ferrers density in the interior to the Hernquist density in the exterior region) is not justified for modeling of the X-ray data. Since the fitted values of $a$ and $n$ depend to some extent on the choice of $a_3$, we examine the effects of $a_3$ varying between the minimum $120''$ and $\infty$. Over this range the best-fit $a$ and $n$ change by less than 10%, where $a_3 = 225''$ yields essentially intermediate parameter values. Upon examination of the three data sets with $a_3$ set to $225''$, we adopt $a = 4''$ and $n = 1.3$. Thus, we model the stellar matter as an oblate spheroid having the density,

$$\rho_{stars} \propto \left[ \left( \frac{4}{225} \right)^2 + m^2 \right]^{-1.3}, \tag{32}$$

where $m$ is the ellipsoidal parameter defined by equation (15). From consideration of the $R$-band isophote shapes we set $\epsilon_{stars} = 1 - q = 0.40$.

In order to completely specify $\rho_{stars}$ we must determine the total stellar mass. Ideally, we would like to assign to $M_{stars}$ the mass associated with the visible stars. Then we could identify $M_{DM}/M_{stars}$ as the ratio of dark matter to stellar matter. Unfortunately, the stellar mass estimates of ellipticals derived from population synthesis techniques (e.g., Pickles 1985; Bacon 1985; Peletier 1989) are very uncertain and are generally modeled to agree with dynamical estimates. Since dynamical masses only yield total masses, the population synthesis estimates of the visible stellar matter may actually contain significant amounts of dark matter. The population synthesis studies generally find that $\Upsilon_B \sim 7\Upsilon_\odot$ for the stellar content of ellipticals, independent of absolute magnitude. Since $B - R$ is essentially constant across NGC 720 (Peletier et al. 1989), the shape of $\rho_{stars}$ in $R$ and $B$ may be assumed equal. Hence, the above mass-to-light ratio translates to a stellar mass $M_{stars} \sim 1.5 \times 10^{11} h_{80} M_\odot$ for NGC 720, where we have used $L_B$ as computed in §4.3.. Note that since this estimate of the stellar mass may contain a significant contribution of dark matter, we may underestimate the mass in dark matter.

## 5.2. X-ray Gas Mass

Neglecting the ellipticity of the gas, the X-ray surface brightness is accurately parameterized by the King function (eq. [3]). By taking $\beta = 0.50$, deprojection of the King function yields the simple expression for the X-ray luminosity density,

$$j_{gas}(r) = \frac{I_0}{2a_X} \left[ 1 + \left( \frac{r}{a_X} \right)^2 \right]^{-3/2}, \tag{33}$$

where $I_0$ is the surface brightness evaluated at $r = 0$ and $a_X = 16''$ is the core radius. We relate $j_{gas}$ to the gas density using equation (9),

$$\rho_{gas} = \left( \frac{\mu^2 m_p^2 I_0}{0.44 a_X \Lambda_{PSPC}(T_{gas})} \right)^{1/2} \left[ 1 + \left( \frac{m a_{3X}}{a_X} \right)^2 \right]^{-3/4}, \tag{34}$$

where we have set $a_{3X} = 375''$ (cf. §2.1.); and in analogy with $\rho_{stars}$ we have expressed $\rho_{gas}$ in terms of the dimensionless ellipsoidal parameter $m$ (cf. eq. [15]). Although we have derived the radial variation of $\rho_{gas}$ assuming spherical symmetry, we set $\epsilon_{gas} = 0.25$ in the models to reflect the shape of the X-ray isophotes. As we show below, the precise form for $\rho_{gas}$ is not particularly important in the models since $M_{gas}/M_{stars}$ is small.

We obtain the mass of the gas by integrating equation (34). For simplicity, and because the isophote shapes are not well constrained for distances greater than $\sim 105''$, we assume spherical symmetry for estimation of $M_{gas}$; this assumption will cause us to overestimate the mass by $\sim 25\%$ if the gas is intrinsically oblate with constant $\epsilon_{gas} = 0.25$ out to $a_{3X}$. Using the single-temperature 90% confidence range for $T_{gas}$ (cf. Table 5), we list in Table 10 values of $M_{gas}$, the volume-averaged particle density ($\overline{n}$), and its associated cooling time ($\overline{\tau}$), all computed within $a_{3X} = 375''$; also listed are the best-fit results for the two-temperature spectrum with solar abundances having emission-weighted temperature 0.74 keV. To facilitate comparison with TFC, we also list these parameters computed within $r = 210''$. TFC, who apply a different technique and assume a 1 keV spectrum with solar abundances, obtain best-fit estimates (scaled to $D = 20$ Mpc) of $M_{gas} = 1.2 \times 10^9 M_\odot$, $\overline{n} = 1.1 \times 10^{-3}$ cm$^{-3}$, and $\overline{\tau} = 2.9 \times 10^9$ yr, in good agreement with our values within TFC's considerable uncertainties.

These values of $M_{gas}$ imply $M_{gas}/M_{stars} \lesssim 1/20$, where $M_{stars}$ was estimated in the previous section; i.e. the influence of the gas on the total potential of the galaxy is negligible. Nevertheless, we included the gas in our models (typically setting $M_{gas}/M_{stars} = 1/50$) and determined that not until $M_{gas}/M_{stars} \sim 1$ does this ratio begin to significantly influence the derived dark matter shapes and masses; i.e. the self-gravity of the gas is not dynamically important.

## 5.3. Results

Having specified $\rho_{stars}$, $\rho_{gas}$, and $M_{gas}/M_{stars}$, the only remaining quantities required to determine the total gravitational potential (eq. [30]) are the dark matter shape parameters $a_3$, $\epsilon_{DM}$, $a_0$, and the mass ratio $M_{DM}/M_{stars}$. For a given value of $M_{DM}/M_{stars}$, the limits on $\epsilon_{DM}$ are obtained in the same manner as in §4.. Because of the stringent constraints placed on temperature gradients by the K-S tests (cf. §4.2.) we restrict ourselves to the isothermal case; any small uncertainties due to temperature gradients will be outweighed by systematic effects resulting from our specific choice of mass models. As a result we employ the isothermal expression (eq. [22]) for $\rho_{gas}$. For simplicity we consider only oblate forms for the stars, gas, and dark matter. In the following section we discuss the position angle offset of the optical and X-ray distributions.

First we examine dark matter having $\rho_{DM} \sim r^{-2}$ and $a_3 = 450''$. In Table 11 we list the results of our fits for several values of $M_{DM}/M_{stars}$. The galaxy without any dark matter is immediately ruled out because $\rho_{stars} + \rho_{gas}$ alone produces an X-ray surface brightness far too steep to account for the data (cf. Figure 10); we mention that polytropic models of the stellar mass will yield acceptable fits to the X-ray surface brightness only for large polytropic indices ($\gamma \sim 1.5$) that imply large temperature gradients that are ruled out by the PSPC spectrum (cf. §2.2.). For $M_{DM}/M_{stars} > 25$, the models highly resemble the single-component case; i.e. the model surface brightness fits the data beautifully and exhibits $\epsilon_{DM}$ ranges virtually identical to $\epsilon_{tot}$ in Table 8. For smaller values of $M_{DM}/M_{stars}$, the fits deteriorate while the $\epsilon_{DM}$ limits remain nearly constant in width but are systematically shifted upwards by $\epsilon \sim 0.02$. However, the core parameter values increase with decreasing mass ratio in order to flatten out the radial profile which is becoming steeper due to the increasing influence of $\rho_{stars}$. By employing the same criteria used in §4.2. for determining the acceptability of fits, we find that for $a_3 = 450''$, $M_{DM}/M_{stars} > 7$ (90% confidence), independent of the distance to the galaxy or the gas temperature; we show in Figure 10 the fit results of a typical model. It then follows that $M_{DM} > 1.1 \times 10^{12} h_{80} M_\odot$ and $M_{tot} > 1.2 \times 10^{12} h_{80} M_\odot$ using the value for $M_{stars}$ adopted in §5.1.. As with the total matter, the fitted parameters do not change substantially over the allowed ranges of $a_3 > a_{3min} = 225''$; e.g., the $\epsilon_{DM}$ limits shift systematically higher by $\sim 0.02$ for $a_3 = 225''$. However, because of the smaller volume the minimum dark mass ratio falls to $M_{DM}/M_{stars} = 4$ for $a_3 = 225''$. We thus conclude that $M_{DM}/M_{stars} \geq 4$ is a firm lower limit, although visual examination of the fits to the radial profile suggest that dark matter at least $\sim 10$ times the stellar mass yields a more accurate description of the data; we mention that the models with temperature gradients give the same results as with the single-component models (cf. §4.2.). The lower limit for the mass derived from $M_{DM}/M_{stars}$ is marginally consistent with the upper end of the confidence interval for the corresponding isothermal $M_{tot}$ in Table 9. This slight discrepancy could be accounted for if the gas is really multi-temperature, or if the galaxy is more distant than 20 Mpc, or if our chosen mass models are not adequate descriptions for the galaxy.

The results for $\rho_{DM}$ having the Hernquist form closely parallel the $\rho_{DM} \sim r^{-2}$ behavior; i.e. for $M_{DM}/M_{stars} > 25$ the total matter results of §4.2. are returned very accurately. Judging by the quality of the fits, the minimum allowed mass ratio for $a_3 = 450''$ is 9 and the $\epsilon_{DM}$ limits change less than $\sim 0.01$ over the range of $M_{DM}/M_{stars}$. For mass ratios decreasing below 25 the values of $a_0$ increase substantially, becoming equal to and exceeding $a_3$ for $M_{DM}/M_{stars} \leq 10$. Such large values for $a_0$ indicate that the $r^{-4}$ regime of the Hernquist density is being suppressed, thus suggesting that an intrinsic profile flatter than the Hernquist form is a more natural description of the dark matter.

We plot in Figure 11(a) the mass of stars, gas, and dark matter as a function of $a = m a_3$ for $\rho_{DM} \sim r^{-2}$ and $a_3 = 450''$ assuming $M_{DM}/M_{stars} = 10$; the plot is normalized to the value of $M_{stars}$ from §5.1.. For comparison we plot in Figure 11(b) $\Upsilon_B(a) = M_{tot}(a)/L_B(a)$, where $M_{tot}(a) = M_{DM}(a) + M_{stars}(a) + M_{gas}(a)$ is the total mass within $a$, and $L_B(a) = \Upsilon_{Bstars} M_{stars}(a)$, where $\Upsilon_{Bstars} \sim 7\Upsilon_\odot$ from §5.1.. The stellar mass dominates the dark matter within $\sim 20''$, but $M_{DM}/M_{stars}$ increases quickly to 1 at $\sim 50''$ corresponding to the optical effective radius ($R_e = 52''$, Burstein et al. 1987). Exterior to $R_e$ the dark matter prevails.

This behavior of $\Upsilon_B$ is consistent with recent optical studies. Both Binney et. al. (1990)

and van der Marel (1991) obtain a nearly constant value for $\Upsilon_B$ in the inner regions of NGC 720; in addition, van der Marel concludes from his study of NGC 720 and 36 other bright ellipticals that $\Upsilon$ is generally not constant in the outermost regions of these ellipticals. This description for $\Upsilon_B(a)$ is consistent with that inferred from other stellar kinematic data (e.g., in de Zeeuw & Franx 1991). More recently, using a two-component model of stars + dark matter, Saglia, Bertin, & Stiavelli (1992; Bertin, Saglia, & Stiavelli 1992) conclude from stellar dynamical analyses of 10 bright round ellipticals, that generally the amount of dark matter inside an effective radius ($R_e$) is of order the stellar mass; typically $\Upsilon_B \sim 7\Upsilon_\odot$ for the stars and $\Upsilon_B \sim 12\Upsilon_\odot$ for the total mass. The analysis of Saglia et al. underestimates the mass if those galaxies, round in projection, are actually flattened along the line-of-sight. Our results for $\Upsilon_B$ obtained by analyzing X-ray data of NGC 720 agree with these studies.

## 6. Discussion

The procedures developed in §4. and §5. to measure the ellipticity of the total gravitating matter and the dark matter both assume a mass ellipsoid of constant shape and orientation. However, the derived shapes are certain only out to distances where the X-ray isophote shapes are well determined. In Figure 12 we illustrate this effect by plotting ellipticity as a function of $a = ma_3$ for the X-ray surface brightness data (cf. Table 3) and a typical single-component matter model. The ellipticity of the data and model show excellent agreement for $a \leq 105''$ with the exception of $a = 60''$; presumably the $a = 60''$ discrepancy is due to the systematic errors discussed in §2.1.2. since the dip is not observed from the results of the isophote fitting. For $a > 105''$, the ellipticities of the model exceed the data which may result from either a real decrease in ellipticity of the gas or to a measurement error due to systematic errors in the computation of $\epsilon_M$ from the data; i.e, the systematic errors discussed in §2.1.2. become more serious as the $S/N$ decreases as does the importance of the background and any other environmental effects. As a result of this uncertainty in the data, our constraints on the shape of the total matter and dark matter are strictly valid only out to $a = 105''$. The minimum acceptable $a_3 = 225''$ is quite insensitive to the relatively small ellipticities of the X-ray isophotes since it is determined from fits to the azimuthally-averaged radial profile. Because the dark matter may be significantly rounder than our models for $a > 105''$ the models may underestimate the total mass by as much as a factor of 2. It would be useful to compare the results for constant shape ellipsoids to models possessing a slow radial variation of ellipticity (e.g., Stäckel potentials); we will explore the effects of different mass models in a future paper. In any event, we must await future missions (i.e. AXAF) with increased sensitivity to obtain precise measurements of the outer X-ray isophotes and thus determine the shape of the dark matter for larger distances.

The misalignment of the projected major axes of the gas and stars is intriguing. We argue in §3.2. that if there were no dark matter, and the stellar ellipsoid is axisymmetric, then the major axes should be aligned. Triaxiality could be the source of such an offset which we will explore in a future paper which will include a ROSAT HRI observation of NGC 720. Another possibility is that the gas and stars are both axisymmetric (e.g., both are oblate) but their axes are not aligned. If indeed the mass in the interior of the galaxy is dominated by the stars as suggested by our models §5.3., we would expect the isophote major axes to gradually align themselves with the stellar matter as the radius decreases. We investigated the effects of such a misalignment for the $M_{DM}/M_{stars} = 7$ models of the previous section. We find that the requirement that the models reproduce the observed position angle offset does not increase the required amount of dark matter. This is simply an effect of the PSPC point spread function smearing out the inner $30''$ where the stellar potential and any corresponding position angle twists become important. The superior resolution of the HRI should enhance our understanding of these issues.

A misalingment of the three-dimensional gas and stellar distributions will also have implications for theories of galaxy formation. In their simulations of hierarchical galaxy formation including gas dynamics, Katz & Gunn (1991) produce objects resembling spiral galaxies where the disk transfers more than 50% of its original angular momentum to the dark halo and forms at an angle of $\sim 30°$. Similar inclinations of the dark halo and stellar matter are observed in related simulations for galaxies of different Hubble types (Neal Katz 1993, private communication).

We have discussed in §3.1. how the interpretation of the shapes of the X-ray isophotes could be clouded if the gas is actually a multi-phase medium. However, it is also possible that in the very center, where the emission from the cold clumps dominate, the shape of the radial profile of the X-ray surface brightness could be distorted by a central peak; e.g., the excess emission due to a cooling flow. With regards to the derived ellipticity of the total matter, our models do not appear to be overly sensitive to the local details of the radial profile. Hence we conclude that the fine details of the state of the gas do not affect the shape determination; the fitted parameters $a_0$ and $\Gamma$ are more sensitive, but typically do not vary by more than $\sim 50\%$.

The shape of the flattened halo we measure for NGC 720 appears to be consistent with standard dissipationless collapse scenarios in a Cold Dark Matter (CDM) universe (Frenk et al. 1988; Katz 1991; Dubinski & Carlberg 1991; Franx, Illingworth, & de Zeeuw 1991; Warren, Quinn, Salmon, & Zurek 1992; cf. Silk & Wyse 1993 for a review). Generally these simulations produce halos which are, on average, flatter than the stellar population with a mean ellipticity $\sim 0.50$. In addition, the simulations of Dubinski & Carlberg (1992) do not produce halos flatter than $\epsilon \sim 0.60$ which happens to be approximately the mean $\epsilon$ of our results. It is also interesting to note that Dubinski & Carlberg (1992) find that their halos are fitted extremely well by a Hernquist density with an extremely small core. These results are also reproduced when dissipation is included in the simulations (Dubinski 1993). This is certainly not true for our models, although the core radii that we derive may be contaminated by the presence of a cooling flow in the innermost region (see above).

Until recently, the evidence for dark matter in normal ellipticals was quite weak (for reviews see Kent 1990; de Zeeuw & Franx 1991; Ashman 1992). Specifically, optical studies of normal ellipticals are generally confined to within $\sim R_e$ where the potential is likely to be dominated by the stars. And the masses of the few galaxies possessing rotation curves calculated from H I emission are uncertain because of uncertainty regarding the shape of the gas orbits. Even the previous X-ray studies of normal ellipticals with *Einstein* (e.g., TFC) have been very uncertain due to the poor constraints on $T(r)$. Recently Saglia et. al. (1993), having obtained accurate velocity dispersions for several ellipticals out to distances greater than $\sim (1-2)R_e$, find strong evidence for dark matter. Maoz & Rix (1993) deduce from observed gravitational lensing statistics that early-type galaxies have dark halos with typical velocity dispersions $\sigma^* > 270$ km s$^{-1}$ for an $L^*$ galaxy. From analysis of the polar ring galaxy NGC 4650, Sackett & Sparke (1991) conclude that there exists a dark matter halo with ellipticity $\sim 0.60$, although with considerable uncertainty. Recent studies of the Galactic halo and the halos of other late-type galaxies show evidence for triaxiality (Franx & de Zeeuw 1992; Kuijken & Tremaine 1993). All of these findings are consistent with our results.

In contrast, analyzing the dynamics of planetary nebulae extending out to $3.5R_e$ in the E0 galaxy NGC 3379, Ciardullo, Jacoby, & Dejonghe (1993) conclude that simple models having a constant mass-to-light ratio fit the data adequately without the need for dark matter. However, they do not demonstrate that dark matter models are inconsistent with their data; e.g., a massive dark matter halo model with anisotropic velocity dispersion. In addition to the possible environmental effects discussed by the authors to explain the "missing" dark matter, the spherical geometry of the stars may represent additional uncertainty. For example, Saglia et al. (1992) employ sophisticated two-component dynamical models to analyze the mass distributions for several bright ellipticals (cf. §5.3.). They caution the reader that their "method seems to underestimate the amount of dark matter present" for intrinsically non-spherical objects seen round in projection. Hence, if NGC 3379 is significantly flattened along the line of sight, Ciardullo et al. likely underestimate the mass of the galaxy. We believe that Ciardullo et al.'s result does not contradict increasing evidence that ellipticals contain large amounts of dark matter.

## 7. Conclusion

We have described (1) a new test for dark matter and alternate theories of gravitation based on the relative geometries of the X-ray and optical surface brightness distributions and an assumed form for the gravitational potential of the optical light, (2) a technique to measure the shapes of the total gravitating matter and dark matter in an ellipsoidal system which is insensitive to the precise value of the temperature of the gas and to modest temperature gradients, and (3) a method to determine the ratio of dark mass to stellar mass (when the self-gravitation of the gas may be ignored) that is dependent on the functional forms for the visible star, gas, and dark mass but independent of the distance to the galaxy or the gas temperature.

We have applied these techniques to X-ray surface brightness data from the ROSAT PSPC of the flattened elliptical galaxy NGC 720. NGC 720 was selected because its flattened stellar distribution ($\epsilon \sim 0.40$) reduces possibilities of significant projection effects, and its large degree of isolation from other large galaxies suggests that the gas is not distorted by environmental effects.

We draw the following conclusions:

1. We compute the ellipticities of the X-ray surface brightness by essentially taking quadrupole moments of the count distribution. The X-ray isophotes are elongated, having $\epsilon \sim 0.25$ for semi-major axis $a \sim 100''$. The major axes of the optical and X-ray isophotes are misaligned by $\sim 30°$

2. The gas does not exhibit either significant radial or azimuthal temperature gradients. A single-temperature ($\sim 0.6$ keV) Raymond-Smith plasma with sub-solar heavy element abundances is a good fit to the data; a two-temperature model (0.5 and 1.1 keV) with solar abundances describes the data just as well.

3. Considering only the relative geometries of the X-ray and optical surface brightness distributions and an assumed form for the potential of the optical light, *we conclude that matter distributed like the optical light cannot produce the observed ellipticities of the X-ray isophotes,* independent of the pressure and temperature of the gas and the value of the stellar mass. This conclusion assumes the conditions of quasi-hydrostatic equilibrium; i.e the shapes of the three-dimensional gas density trace the three-dimensional gravitational potential. We discuss the viability of this assumption in §3.1.. Since this analysis is confined to the region where Milgrom's Modification of Newtonian Dynamics (MOND) predicts Newton's laws to apply, *we conclude that MOND does not eliminate the need for dark matter in NGC 720.*

4. Employing essentially the technique of Buote & Canizares (1992; Buote 1992) we use the *shape* of the X-ray surface brightness to constrain the *shape* of the total gravitating matter. The total matter is modeled as an oblate or prolate spheroid of constant shape and orientation having either a Ferrers ($\rho \sim r^{-n}$) or Hernquist density. Assuming the X-ray gas is in hydrostatic equilibrium with the potential generated by this mass, we construct a model X-ray gas distribution.

5. We determine the ellipticity of the total gravitating matter to be $\epsilon \sim 0.50 - 0.70$. Using the single-temperature model we estimate a total mass $(0.41 - 1.4) \times 10^{12} h_{80} M_\odot$ interior to a spheroid having semi-major axis ranging from $21.8 - 43.6 h_{80}$ kpc. Ferrers densities as steep as $r^{-3}$ do not fit the data, but the $r^{-2}$ and Hernquist models yield excellent fits.

6. We estimate the mass distributions of the stars and the gas by deprojecting their observed major-axis surface brightness profiles. We then fit the dark matter directly and find shapes in good agreement with those derived for the total matter. These fits yield a distance-independent and temperature-independent measurement of the ratio of dark mass to stellar mass $M_{DM}/M_{stars}$, but it is dependent on the models assumed for the three mass components of the galaxy. We estimate at minimum $M_{DM}/M_{stars} \geq 4$ interior to a spheroid of semi-major axis $21.8 h_{80}$ kpc corresponding to a total mass $(8.0 \times 10^{11} h_{80} M_\odot)$ slightly greater than that derived from the single-temperature models at $D = 20h_{80}$ Mpc $(4.1 - 7.5 \times 10^{11} h_{80} M_\odot)$. More plausible values are $M_{DM}/M_{stars} \sim 10$ out to $\sim 30h_{80}$ Mpc. The estimates for $M_{DM}/M_{stars}$ may be lower than in reality since $M_{stars}$ may contain a significant portion of dark matter.

Similar studies need to be performed on other galaxies in various environments to determine whether a flattened halo is a general property of ellipticals. In addition, the new proposed test for dark matter and alternate theories of gravitation needs to be applied to other galaxies (and perhaps clusters of galaxies) in order to ascertain the generality of our conclusions regarding MOND.

We are indebted to Paul Schechter and James Binney for suggesting the importance of studying X-ray properties of flattened ellipticals. Special thanks are given to an anonymous referee for his/her careful reading of the manuscript and whose many suggestions greatly improved the content and presentation of the paper. We express sincere gratitude to James Binney, Mordehai Milgrom, and Scott Tremaine for their careful reading of the manuscript and for several valid criticisms, particularly raising the issue of gas rotation. We would also like to thank Marijn Franx for clarifying aspects of his Tully-Fisher relation for ellipticals. We gratefully acknowledge Christopher Becker, Edmund Bertschinger, Giuseppina Fabbiano, Eric Gaidos, Neal Katz, Eugene Magnier, and Paul Schechter for insightful discussions, Jonathon Woo for assistance with XSPEC, and the friendly people of hotseat@cfa.harvard.edu (especially Kathy Manning) who cheerfully answered our many questions regarding PROS. We thank John Tonry for providing us with his $\chi^2$ minimization programs. We also acknowledge use of the SIMBAD data base. Supported in part by NASA grant NAG5-1656, NASGW-2681 (through subcontract SVSV2-62002 from Smithsonian Astrophysical Observatory), and NAS8-38249.

## A. Analytical Calculation of $\Delta\epsilon_M$ and $\Delta\theta_M$

We derive the statistical uncertainty of the ellipticity and position angle for the iterative moment technique described in §2.1.2.. Recall that $\epsilon_M$ and $\theta_M$ are complicated functions of the moments $\mu_{mn}$ (eq. [4]) which are themselves weighted averages over the whole aperture. Since the photon fluctuations from pixel to pixel are uncorrelated, we have for the variance in ellipticity, $(\Delta\epsilon_M)^2$, and position angle, $(\Delta\theta_M)^2$,

$$(\Delta\epsilon_M)^2 = \sum_{i=1}^{P} \left(\frac{\partial\epsilon_M}{\partial n_i}\right)^2 \sigma_{n_i}^2 \quad \text{and} \quad (\Delta\theta_M)^2 = \sum_{i=1}^{P} \left(\frac{\partial\theta_M}{\partial n_i}\right)^2 \sigma_{n_i}^2, \tag{A1}$$

where $N = \sum_{i=1}^{P} n_i$ is the total number of pixels in the aperture considered, and $\sigma_{n_i}^2$ is the variance of the counts, $n_i$, in the $i$'th pixel: for Poisson statistics, $\sigma_{n_i}^2 = n_i$. We begin by expressing $\epsilon_M$ (eq.

[5]) and $\theta_M$ (eq. [6]) in terms of the moments $\mu_{mn}$. Solving equation (7) for $\Lambda_\pm$ yields,

$$\Lambda_\pm = \left( \frac{-b \pm \sqrt{b^2 - 4c}}{2} \right)^{1/2}, \tag{A2}$$

where $b = -(\mu_{02} + \mu_{20})$ and $c = \mu_{02}\mu_{20} - \mu_{11}^2$. The moments from equation (4) take the explicit form:

$$\mu_{02} = \frac{1}{N} \sum_i n_i Y_i^2 - \left( \frac{1}{N} \sum_i n_i Y_i \right)^2, \tag{A3}$$

$$\mu_{20} = \frac{1}{N} \sum_i n_i X_i^2 - \left( \frac{1}{N} \sum_i n_i X_i \right)^2, \tag{A4}$$

$$\mu_{11} = \frac{1}{N} \sum_i n_i X_i Y_i - \frac{1}{N^2} \sum_i n_i X_i \sum_i n_i Y_i, \tag{A5}$$

where we have suppressed the upper limit, $P$, on the summations in the interest of compact notation. In practice we set $X_i \equiv x_i - \bar{x}_i$ and $Y_i \equiv y_i - \bar{y}_i$; in the following we will neglect the derivatives of the additional centroid terms since they contribute terms that are of order $\frac{1}{N}$ with respect to the other undifferentiated terms. Since N is a large number ($> 100$) for all our apertures, we may safely neglect this contribution.

Substituting $\Lambda_\pm$ into the expressions for $\epsilon_M$ and $\theta_M$ and taking the derivative with respect to $n_i$ gives,

$$\frac{\partial \epsilon_M}{\partial n_i} = \left( b + \sqrt{b^2 - 4c} \right) \left( \frac{1}{4c^{3/2}} \frac{\partial c}{\partial n_i} \right) - \frac{1}{2\sqrt{c}} \left( \frac{\partial b}{\partial n_i} + \frac{b \partial b / \partial n_i - 2 \partial c / \partial n_i}{\sqrt{b^2 - 4c}} \right), \tag{A6}$$

$$\frac{\partial \theta_M}{\partial n_i} = \left[ 1 + \left( \frac{\mu_{11}}{\Lambda_+^2 - \mu_{02}} \right)^2 \right]^{-1} \left\{ \frac{\partial \mu_{11} / \partial n_i}{\Lambda_+^2 - \mu_{02}} - \frac{\mu_{11}}{(\Lambda_+^2 - \mu_{02})^2} \left[ 2\Lambda_+ \frac{\partial \Lambda_+}{\partial n_i} - \frac{\partial \mu_{02}}{\partial n_i} \right] \right\}, \tag{A7}$$

where

$$\frac{\partial \Lambda_+}{\partial n_i} = 2^{-3/2} \left( -b + \sqrt{b^2 - 4c} \right)^{-1/2} \left[ -\frac{\partial b}{\partial n_i} + \frac{b \partial b / \partial n_i - 2 \partial c / \partial n_i}{\sqrt{b^2 - 4c}} \right], \tag{A8}$$

where the derivatives of $b$ and $c$ follow straightforwardly from their above definitions. All that now remains is to compute the derivatives of the moments. Keeping terms only to order $1/N$ we find:

$$\frac{\partial \mu_{02}}{\partial n_i} = \frac{1}{N} Y_i^2 - \frac{2}{N} Y_i \left( \frac{1}{N} \sum_k n_k Y_k \right), \tag{A9}$$

$$\frac{\partial \mu_{20}}{\partial n_i} = \frac{1}{N} X_i^2 - \frac{2}{N} X_i \left( \frac{1}{N} \sum_k n_k X_k \right), \tag{A10}$$

$$\frac{\partial \mu_{11}}{\partial n_i} = \frac{1}{N} X_i Y_i - \frac{1}{N} X_i \left( \frac{1}{N} \sum_k n_k X_k \right) - \frac{1}{N} Y_i \left( \frac{1}{N} \sum_k n_k Y_k \right). \tag{A11}$$

By substituting the expressions for $\partial \epsilon_M / \partial n_i$ and $\partial \theta_M / \partial n_i$ into equation (A1) , one obtains the 68% confidence statistical uncertainties $\Delta \epsilon_M$ and $\Delta \theta_M$. Multiplying these 68% errors by $\sqrt{2.71}$ gives 90% error estimates. We have verified the reliability of these uncertainty estimates through the Monte Carlo simulations described in §2.1.2..

## B. Projections of Non-Similar Spheroids

There is a paucity of simple, yet flexible, analytic models for non-similar spheroids. By flexible we mean that the models extant in the literature generally do not allow one to easily impose a specific ellipticity function ($\epsilon(r), r = \sqrt{x^2 + y^2 + z^2}$) on the model; e.g. Stäckel models (e.g., Dejonghe & de Zeeuw 1988), models consisting of a multipole decomposition into monopole and quadrupole terms (e.g., Kochanek 1991), and models constructed by adding individual homoeoids of varying axial ratio (Schramm 1994). In addition, the first two of these models are not exactly spheroidal. In order to achieve the desired flexibility, we prefer to generalize the similar spheroid case by considering functions stratified on surfaces of constant

$$\xi^2 = x^2 + y^2 + z^2/q^2, \tag{B12}$$

where $q = q(r)$ is the radially-varying axial ratio. These surfaces, like the previously mentioned examples, are not true spheroids. However they are good approximations to spheroids for reasonable $q(r)$, their deviations being characterized by slight "boxyness". Hence, in these models $q < 1$ corresponds to an oblate pseudo spheroid, $q > 1$ corresponds to a prolate pseudo spheroid.

We are ultimately interested in functions that represent the X-ray gas volume emissivities and gravitational potentials of elliptical galaxies. It follows that we may restrict ourselves to functions whose radial dependence is not flatter than $\log \xi$ (corresponding to flattest reasonable potentials) and not steeper than $\xi^{-4}$ (corresponding to volume densities appropriate to the outer regions of a de Vaucouleurs Law). We also demand that our functions possess ellipticity ($\epsilon = 1 - q$) gradients that are typical of assumed potentials in elliptical galaxies (cf. Figure 2-13 in Binney & Tremaine 1987). Such ellipticities are smooth and monotonically decreasing, and have central ellipticity no greater than $\epsilon \sim 0.40$ corresponding to E6 galaxies.

A simple parametrization of the ellipticity of the pseudo spheroids that qualitatively obeys these restrictions is given conveniently by,

$$\epsilon(r) = \frac{2\epsilon_c}{1 + r/r_c}, \tag{B13}$$

where $\epsilon_c = \epsilon(r_c) = \epsilon(0)/2$. For $r \gg r_c$, $\epsilon(r) \sim r^{-1}$ which is somewhat steeper than the gradients of the assumed theoretical potentials. We desire this behavior since our intent is to study the effects of ellipticity gradients of a three dimensional distribution on the ellipticities of the contours of its projection. The projections of the pseudo spheroids with $\epsilon(r)$ should exhibit the maximum deviations from the similar spheroid case expected of ellipticity gradients consistent with the above restrictions.

We consider the functions $\log(a_0^2 + \xi^2)$ and $(a_0^2 + \xi^2)^{-2}$ with $q(r) = 1 - \epsilon(r)$ as given above, where $a_0$ is the core parameter. We assume $a_0$ is the same for both functions because the core parameter should be very similar for the potential and X-ray emissivity under the conditions of hydrostatic equilibrium and reasonably small temperature gradients (eq. [1]). Moreover, for steep (negative) temperature gradients $a_0$ of the mass (and potential) may be significantly smaller than that of the gas. This has the effect of steepening the radial slope of the inner part of the potential, thus bringing the radial slope into slightly better agreement with the steeper gas emissivity; i.e. steep negative temperature gradients will give smaller core parameters for the potential that yield projected ellipticity deviations smaller than in the isothermal case. We set $\epsilon_c = 0.20$ so the model potentials will include the flattest potentials expected for ellipticals; also, smaller values of $\epsilon_c$ reflect more spherical objects whose axial ratios are less sensitive to projection. There are three distinct regimes that characterize the behavior of these pseudo spheroids: (1) $a_0 \gg r_c$, (2) $a_0 \sim r_c$, and (3) $a_0 \ll r_c$. In Figure 13 we plot the projections of these functions (edge-on) in each regime; i.e. $a_0 = r_c/10$, $a_0 = r_c$, and $a_0 = 10 r_c$. The figures show a consistent picture of the difference in ellipticity ($\Delta\epsilon$) of the logarithmic and $r^{-4}$ projections. First, the logarithmic function projects to contours that are noticeably rounder than the $r^{-4}$ and the three dimensional ellipticity. This effect arises because contributions from the rounder, outermost three dimensional surfaces to the projection are more important to the flat logarithmic function than the steep $r^{-4}$ model. We also observe the anticipated correlation between the magnitude of $\Delta\epsilon$ and the gradient in $\epsilon(r)$; i.e. the steeper the gradient in ellipticity, the larger is $\Delta\epsilon$. However, in all regimes $\Delta\epsilon \lesssim 0.04$ for $a > a_0$ and never exceeds 0.06 for all $a$; these results are identical for the oblate and prolate pseudo spheroids. Thus, assuming quasi-hydrostatic equilibrium in elliptical galaxies, *the shapes of the X-ray isophotes and the projected potentials are approximately the same, with maximum deviations of $\Delta\epsilon \lesssim 0.04$ outside of the core region.* For two functions not having such disparate radial slopes and/or flatter ellipticity profiles, the discrepancy in projected ellipticities will be significantly smaller. We illustrate this point with a concrete example applied to NGC 720.

Suppose the only significant mass component in NGC 720 is that due to the visible stars. Then quasi-hydrostatic equilibrium requires that the three dimensional X-ray gas emissivity ($j_{gas}$) has the same three dimensional contours as the gravitational potential generated by the visible stars ($\Phi_{stars}$). As we discuss in §5.1., the visible star density is reasonably approximated by an oblate spheroid with radial dependence $\rho_{stars} \sim r^{-2.6}$ and $\epsilon = 0.40$. This density yields a potential that approximately behaves as $\Phi_{stars} \sim r^{-0.6}$; its contours are moderately flattened at the center ($\epsilon \sim 0.20$) and become monotonically rounder with distance. In §5.2. we show that $j_{gas} \sim r^{-3}$, with a core parameter $a_0 = 16''$.

We now examine the projections of $\Phi_{stars}$ and $j_{gas}$. We parametrize the emissivity as a pseudo spheroid $j_{gas} \propto (a_0^2 + \xi^2)^{-3/2}$, where we assign the $\epsilon(r)$ associated with $\Phi_{stars}$. In order to make a consistent comparison, we also employ the pseudo spheroid construction for $\Phi_{stars}$. That is, after computing $\Phi_{stars}$ numerically, we fit the ellipticity profile ($\epsilon(r)$) along the major axis. An acceptable fit is obtained using a function consisting of products of equation (B13) with additional parameters: $\epsilon(r) \propto (1 + (r/r_c)^a)^{-1}(1 + (r/r_d)^b)^{-1}$, where $r_c$, $r_d$, $a$, and $b$ are free parameters. In Figure 14 (a) we plot this fitted $\epsilon(r)$ and the exact ellipticity profile of $\Phi_{stars}$ obtained numerically. The fitted function yields a good qualitative representation of the three dimensional potential ellipticity.

Using this $\epsilon(r)$, we then construct $\Phi_{stars} \propto (a_0^2 + \xi^2)^{-0.3}$, where $a_0 = 16''$; note the results are not sensitive to the precise choice of $a_0$ (see above). In Figure 14 (b) we show the results of the projections for oblate spheroids (the prolate case gives the same qualitative results as is expected since we are dealing with relatively small ellipticities); note that the integration is performed only within a spheroid having a major axis of $400''$, that being the extent of the X-ray gas. For comparison we plot in Figure 14 (a) the projection of $\Phi_{stars}$ obtained from direct numerical calculation. Notice that our approximation to $\epsilon(r)$ is slightly steeper than the exact case and that the exact projected ellipticities deviate less from the three dimensional ellipticities than for the pseudo spheroid case because the exact case is more closely related to a similar spheroid. As for the pseudo spheroids, the agreement between the projected potential and the projected emissivity is excellent, the maximum deviation being $\Delta\epsilon \sim 0.02$. This value is less than the statistical uncertainty in the measured values of ellipticity of the X-ray isophotes (§2.1.2.). Thus quasi-hydrostatic equilibrium implies that *if the stars are the dominant contribution to the gravitational potential in NGC 720, then the PSPC X-ray isophotes and projected potential contours have virtually identical shapes.*

Table 1: ROSAT Observation of NGC 720

| ROSAT Sequence No. | Date Observed | R.A.;[a] Dec | R.A.[b] Dec | Exposure Time | Flux[c] (erg cm$^{-2}$ s$^{-1}$) |
|---|---|---|---|---|---|
| rp600005 | January, 1992 | $1^h53^m00^s.4$ $-13°44'18''$ | $1^h53^m00^s.0$; $-13°44'20''$ | 23108 sec | $9.76 \times 10^{-13}$ $8.52 \times 10^{-13}$ |

[a]Optical center from Dressler, Schechter, & Rose (1986) precessed to J2000 coordinates.

[b]X-ray centroid (J2000) computed in this paper.

[c]Computed in 400'' radius circle for energy range 0.2 - 2.4 keV; 0.4 - 2.4 keV.

Table 2: Fits to King Function

| Observers | Best Fit $a_x$ (arcsec) | 90% Range | Best Fit $\beta$ | 90% Range | $\chi^2_{min}$ | Degrees of Freedom | R (arcsec) |
|---|---|---|---|---|---|---|---|
| This Paper | 16.0 | 12.0 - 20.7 | 0.51 | 0.49 - 0.53 | 15.6 | 22 | 375 |
| TFC* | 3 | < 37 | 0.45 | 0.40 - 0.50 | 12.9 | 8 | 495 |

*Energy range 0.2 - 4 keV.

Table 3: X-ray Ellipticities

| $a$[a] | $r$[b] | $\epsilon_M$[c] | $\Delta\epsilon_M$[c] | Counts[d] | $a_{in}$[e] | $r$[b] | $\epsilon_M$[f] | $\Delta\epsilon_M$[f] | Counts[d] | $r$[b] | $\epsilon_{iso}$[g] | $\Delta\epsilon_{iso}$[g] | $\epsilon_{opt}$[h] |
|---|---|---|---|---|---|---|---|---|---|---|---|---|---|
| 30 | 29 | 0.08 | 0.06 | 476 | | | | | | 29 | 0.08 | 0.08 | 0.42 |
| 45 | 42 | 0.13 | 0.05 | 689 | | | | | | 41 | 0.16 | 0.09 | 0.44 |
| 60 | 57 | 0.09 | 0.05 | 871 | 30 | 57 | 0.11 | 0.06 | 398 | 54 | 0.18 | 0.08 | 0.48 |
| 75 | 67 | 0.20 | 0.05 | 953 | 45 | 66 | 0.23 | 0.06 | 306 | 65 | 0.24 | 0.14 | 0.46 |
| 90 | 78 | 0.25 | 0.05 | 1011 | 60 | 74 | 0.32 | 0.09 | 229 | 72 | 0.36 | 0.09 | 0.46 |
| 105 | 91 | 0.25 | 0.05 | 1090 | 75 | 87 | 0.32 | 0.09 | 146 | 93 | 0.22 | 0.14 | 0.44 |
| 120 | 112 | 0.13 | 0.06 | 1184 | | | | | | | | | 0.44 |
| 135 | 124 | 0.15 | 0.07 | 1219 | | | | | | | | | |
| 150 | 137 | 0.16 | 0.08 | 1223 | 105 | 140 | 0.13[i] | 0.07[i] | 279[i] | | | | |
| 225 | 201 | 0.20[i] | 0.05[i] | 1774[i] | 150 | 200 | 0.21[i] | 0.09[i] | 251[i] | | | | |

[a]Semi-major axis of aperture in arcseconds.

[b]Effective radius of aperture $r = (ab)^{1/2}$ (arcseconds), where $b = (1 - \epsilon)a$.

[c]Computed with an elliptical aperture containing all counts interior to $a$ (5″ pixels); $\Delta\epsilon_M$ represents 90% confidence statistical uncertainties.

[d]Counts interior to aperture $(0.4 - 2.4)$ keV.

[e]Inner semi-major axis of annular aperture in arcseconds.

[f]Computed with elliptical annular aperture between $a$ and $a_{in}$ (5″ pixels); $\Delta\epsilon_M$ represents 90% confidence statistical uncertainties.

[g]Results from fitting ellipses to the X-ray isophotes; $\Delta\epsilon_{iso}$ represents 68% confidence statistical uncertainties.

[h]$R-$band optical ellipticities taken from Peletier et al. (1989).

[i]Computed from image with 15″ pixels.

Table 4: X-ray Position Angles (N through E)

| $a^a$ | $r^b$ | $\theta_M{}^c$ | $\Delta\theta_M{}^c$ | $a_{in}{}^d$ | $r^b$ | $\theta_M{}^e$ | $\Delta\theta_M{}^e$ | $r^b$ | $\theta_{iso}{}^f$ | $\Delta\theta_{iso}{}^f$ | $\theta_{opt}{}^g$ |
|---|---|---|---|---|---|---|---|---|---|---|---|
| 30 | 29 | 83 | 24 | | | | | 29 | 119 | 28 | 142 |
| 45 | 42 | 125 | 12 | | | | | 41 | 126 | 18 | 142 |
| 60 | 57 | 118 | 17 | 30 | 57 | 116 | 17 | 54 | 109 | 14 | 141 |
| 75 | 67 | 111 | 7 | 45 | 66 | 106 | 9 | 59 | 112 | 19 | 142 |
| 90 | 78 | 113 | 6 | 60 | 74 | 116 | 8 | 73 | 115 | 9 | 144 |
| 105 | 91 | 116 | 6 | 75 | 87 | 120 | 10 | 93 | 112 | 21 | 144 |
| 120 | 112 | 117 | 15 | | | | | | | | 136 |
| 135 | 124 | 102 | 14 | | | | | | | | |
| 150 | 137 | 102 | 15 | 105 | 140 | 104[h] | 18[h] | | | | |
| 225 | 201 | 107[h] | 8[h] | 150 | 200 | 124[h] | 15[h] | | | | |

[a] Semi-major axis of aperture in arcseconds

[b] Effective radius of aperture $r = (ab)^{1/2}$, where $b = (1 - \epsilon)a$.

[c] Position angle (degrees) computed with an elliptical aperture containing all counts interior to $a$ (5″ pixels); $\Delta\theta_M$ represents 90% confidence statistical uncertainties.

[d] Inner semi-major axis of annular aperture in arcseconds.

[e] Position angle computed with elliptical annular aperture between $a$ and $a_{in}$ (5″ pixels); $\Delta\theta_M$ represents 90% confidence statistical uncertainties.

[f] Results from fitting ellipses to the X-ray isophotes; position angle in degrees and $\Delta\theta_{iso}$ represents 68% confidence statistical uncertainties.

[g] $R-$band optical position angles (degrees) taken from Peletier et al. (1989).

[h] Computed from image with 15″ pixels.

Table 5: Spectral Data and Fit Results

| Region | Model | $\chi^2_{min}$ | dof* | $N_H$ cm$^{-2}$ | Abun (% solar) | $T$ (keV) |
|---|---|---|---|---|---|---|
| 0″- 400″ | $1T^{\dagger}$ | 25.1 | 25 | $(0.1 - 3.2) \times 10^{20}$ | 8 - 60 | 0.48 - 0.69 |
| 0″- 400″ | $2T^{\ddagger}$ | 24.0 | 24 | $4 \times 10^{19}$ | 100 | (0.44, 1.1) |
| 0″- 60″ | $1T$ | 21.3 | 18 | $(0.5 - 2) \times 10^{20}$ | 10 - 40 | 0.5 - 0.7 |
| 120″- 400″ | $1T$ | 10.2 | 12 | $< 4 \times 10^{20}$ | 1 - 80 | 0.4 - 0.8 |
| (A) | $1T$ | 15.6 | 16 | $< 2 \times 10^{20}$ | 10 - 80 | 0.5 - 0.7 |
| (B) | $1T$ | 20.0 | 17 | $(1 - 4) \times 10^{20}$ | 5 - 25 | 0.5 - 0.7 |

*Degrees of freedom. The energy ranges are: (0.2 - 2.4) keV for 0″- 400″; (0.2 - 1.7) keV for 0″- 60″; (0.2 - 0.28, 0.4 - 1.4) keV for 120″- 400″; (0.2 - 0.37, 0.4 - 1.6) keV for (A); and (0.2 - 1.5, 1.6 - 1.7) keV for (B).

$\dagger$ Single-temperature Raymond-Smith model. 90% confidence estimates for parameters are shown for 0″- 400″, 68% confidence for the others.

$\ddagger$ Two-temperature Raymond-Smith model with abundances fixed at 100% solar. Only the best-fit values are displayed.

Table 6: Oblate Stellar Equipotential Ellipticities

| $a$ (arcsec) | $\epsilon_x$ | $\epsilon_{pot}$ 3-D | $\epsilon_{pot}$ 2-D | $\epsilon_{pot}$ $\langle$2-D$\rangle_{PSPC}$ | $\epsilon_{isophote}$ $\langle$2-D$\rangle_{PSPC}$ |
|---|---|---|---|---|---|
| 30 | 0.02 - 0.14 | 0.13 | 0.13 | 0.11 | 0.06 |
| 45 | 0.08 - 0.18 | 0.12 | 0.12 | 0.11 | 0.09 |
| 60 | 0.04 - 0.14 | 0.11 | 0.11 | 0.11 | 0.10 |
| 75 | 0.15 - 0.25 | 0.11 | 0.11 | 0.10 | 0.10 |
| 90 | 0.20 - 0.30 | 0.11 | 0.10 | 0.10 | 0.10 |
| 105 | 0.20 - 0.30 | 0.10 | 0.10 | 0.10 | 0.10 |
| 120 | 0.07 - 0.19 | 0.10 | 0.10 | 0.09 | 0.10 |
| 135 | 0.08 - 0.22 | 0.10 | 0.09 | 0.09 | 0.09 |
| 150 | 0.08 - 0.24 | 0.10 | 0.09 | 0.09 | 0.09 |
| 225 | 0.15 - 0.25 | 0.07 | 0.05 | 0.05 | 0.05 |

Note. — Ellipticities listed as a function of semi-major axis $a$. $\epsilon_x$ are the 90% limits for $\epsilon_M$ computed for the X-ray image in Table 3. $\epsilon_{pot}$ is the ellipticity of the stellar isopotentials in three dimensions (3-D), 2-D (i.e. projected along the line of sight), and 2-D convolved with the PSF of the PSPC. $\epsilon_{isophote}$ is the expected ellipticity of the isophotes if the gas is an isothermal ideal gas (i.e. contours of constant projected $\rho_{gas}^2$; cf. Table 11).

Table 7: Monte Carlo Ellipticities

| $a$ | Oblate | | | Prolate | | |
|---|---|---|---|---|---|---|
| (arcsec) | $\bar{\epsilon}_M$ | 90% | 99% | $\bar{\epsilon}_M$ | 90% | 99% |
| 30 | 0.07 | 0.00 - 0.13 | 0.00 - 0.24 | 0.10 | 0.00 - 0.22 | 0.00 - 0.37 |
| 45 | 0.08 | 0.00 - 0.15 | 0.00 - 0.20 | 0.10 | 0.00 - 0.19 | 0.00 - 0.29 |
| 60 | 0.09 | 0.00 - 0.15 | 0.00 - 0.22 | 0.10 | 0.00 - 0.19 | 0.00 - 0.27 |
| 75 | 0.10 | 0.01 - 0.17 | 0.00 - 0.23 | 0.11 | 0.01 - 0.19 | 0.00 - 0.25 |
| 90 | 0.10 | 0.01 - 0.18 | 0.00 - 0.24 | 0.11 | 0.01 - 0.19 | 0.00 - 0.25 |
| 105 | 0.10 | 0.00 - 0.19 | 0.00 - 0.25 | 0.11 | 0.01 - 0.20 | 0.00 - 0.25 |
| 120 | 0.11 | 0.01 - 0.20 | 0.00 - 0.26 | 0.11 | 0.01 - 0.20 | 0.00 - 0.25 |
| 135 | 0.11 | 0.01 - 0.20 | 0.00 - 0.26 | 0.11 | 0.01 - 0.20 | 0.00 - 0.26 |

Note. — Results of 1000 Monte Carlo simulations of the constant mass-to-light model. Listed are ellipticities as a function of semi-major axis $a$ computed using the iterative moment technique described in §2.1.2.. $\bar{\epsilon}_M$ is the mean value of $\epsilon_M$ for all the simulations and 90% and 99% are the corresponding confidence limits

Table 8: Total Gravitating Matter Shape Results ($\epsilon_{tot}$)

| Density Model | Oblate $\epsilon_{tot}$ | Prolate $\epsilon_{tot}$ | $\chi^2_{min}$ [a] | $a_0$ (arcsec)[b] | $|\Gamma|$[b] |
|---|---|---|---|---|---|
| $\rho \sim r^{-2}$ | 0.52 - 0.74 | 0.49 - 0.65 | 15 | 6.9 - 14.9 | 5.63 - 6.14 |
| $\rho \sim r^{-3}$ | 0.50 - 0.72 | 0.45 - 0.62 | 35 | 32.4 - 49.0 | 4.83 - 5.09 |
| Hernquist | 0.50 - 0.71 | 0.47 - 0.63 | 20 | 133 - 213 | 5.43 - 5.80 |

[a]Typical minimum $\chi^2$ (22 dof) for $\epsilon_{tot}$ ranges in columns 2 and 3.

[b]90% confidence values for oblate $\epsilon_{tot}$ range in column 2.

Table 9: Total Spheroidal Mass[†]

| Density Model | $a_3$[‡] | Oblate | | Prolate | |
|---|---|---|---|---|---|
| | | $M_{tot}(10^{12}M_\odot)$ | $\Upsilon_B(\Upsilon_\odot)$ | $M_{tot}(10^{12}M_\odot)$ | $\Upsilon_B(\Upsilon_\odot)$ |
| $\rho \sim r^{-2}$ | 450 | 0.79 - 1.4 | 35.9 - 62.7 | 0.61 - 1.1 | 27.7 - 51.8 |
| $\rho \sim r^{-2}$ | $(225, 260)^*$ | 0.45 - 0.75 | 20.6 - 34.3 | 0.41 - 0.73 | 18.8 - 33.2 |
| Hernquist | 450 | 0.64 - 1.1 | 28.9 - 51.8 | 0.54 - 1.0 | 24.6 - 45.9 |

[†]Assuming $D = 20h_{80}$ Mpc.

[‡]Semi-major axis in arcseconds ($10'' \sim 1$ kpc).

[*](oblate, prolate).

Table 10: Total X-ray Gas Mass[†]

| Model[‡] | $M_{gas}$ $(10^9 h_{80}^{5/2} M_\odot)$ | | $\overline{n}$ $(10^{-3} h_{80}^{-1/2}$ cm$^{-3})$ | | $\overline{\tau}$ $(10^9 h_{80}^{1/2}$ yr$)$ | |
|---|---|---|---|---|---|---|
| | $r = 210''$ | $r = 375''$ | $r = 210''$ | $r = 375''$ | $r = 210''$ | $r = 375''$ |
| $1T$ | 2.1 - 3.0 | 5.9 - 8.2 | 2.5 - 3.6 | 0.99 - 1.36 | 2.8 - 2.9 | 7.1 - 7.3 |
| $2T^*$ | 1.5 | 4.0 | 1.66 | 0.65 | 2.0 | 5.2 |

[†]0.4 - 2.4 keV.

[‡]Consult the spectral models in Table 5.

[*]We have used the emission-weighted temperature 0.74 keV.

Table 11: Dark Matter Shape Results[†]

| $\frac{M_{DM}}{M_{stars}}$ | $\epsilon_{DM}$ | $\chi^2_{min}$ [a] | $a_0$ (arcsec)[b] | $|\Gamma|$[b] |
|---|---|---|---|---|
| 100 | 0.52 - 0.74 | 14.7 - 15.5 | 7.7 - 16.4 | 5.65 - 6.14 |
| 50 | 0.52 - 0.74 | 14.7 - 15.5 | 8.7 - 19.4 | 5.67 - 6.15 |
| 25 | 0.52 - 0.75 | 14.7 - 15.4 | 11.0 - 23.1 | 5.71 - 6.20 |
| 10 | 0.54 - 0.76 | 15.8 - 16.4 | 23.8 - 65.1 | 5.92 - 6.70 |
| 8 | 0.56 - 0.77 | 20.4 - 24.5 | 30.3 - 92.1 | 6.02 - 6.63 |
| 7 | 0.57 - 0.78 | 28.7 - 35.8 | 33.1 - 93.9 | 6.01 - 6.51 |
| 0[c] | ... | 605 | ... | 5.2 |

[†]Oblate dark matter model with density, $\rho_{DM} \sim r^{-2}$ and $a_3 = 450''$.

[a]22 degrees of freedom.

[b]90% confidence values over $\epsilon_{DM}$ interval in column 2.

[c]Only the best-fit values are listed.

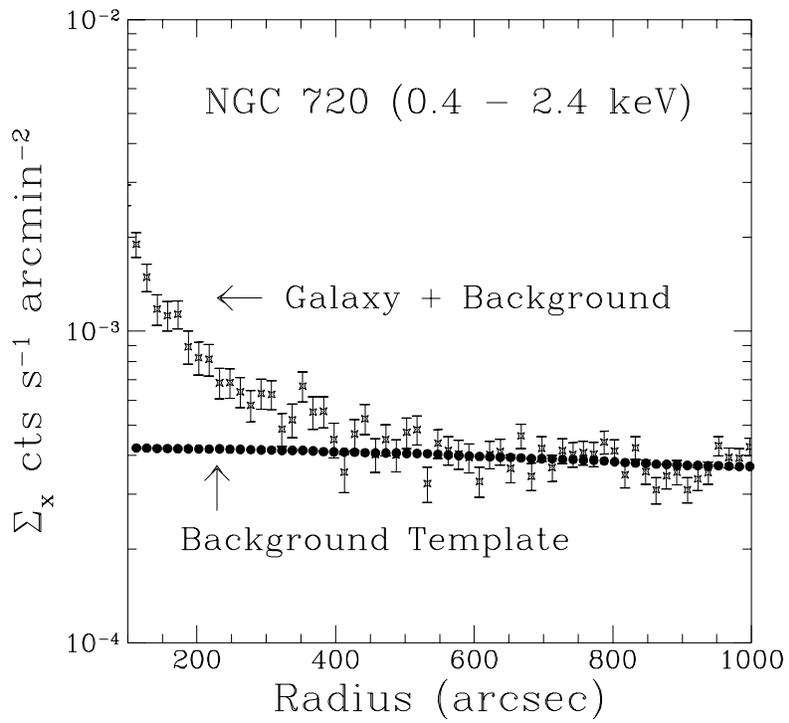

Fig. 1.—
Azimuthally-averaged radial profile (15″ bins) of the image and the background template both corrected for the effects of exposure variations, vignetting, and embedded point sources.

Fig. 2.—
Contour map of the X-ray surface brightness of the elliptical galaxy NGC 720; the contours are separated by a factor of 2 in intensity. The image has been corrected for the effects of exposure variations, vignetting, embedded point sources, and background; the point sources have simply been removed from the image thus causing some of the apparent asymmetries for radii greater than about $150''$; e.g. the isolated contour in the upper right. The image has been smoothed for visual clarity with a Gaussian of $\sigma = 11.25''$, although the image used for analysis is not smoothed in any manner.

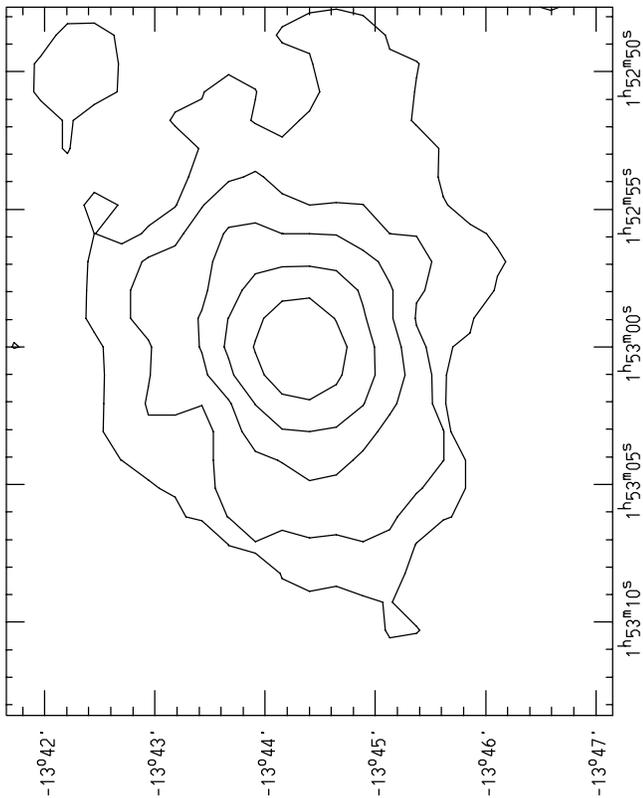

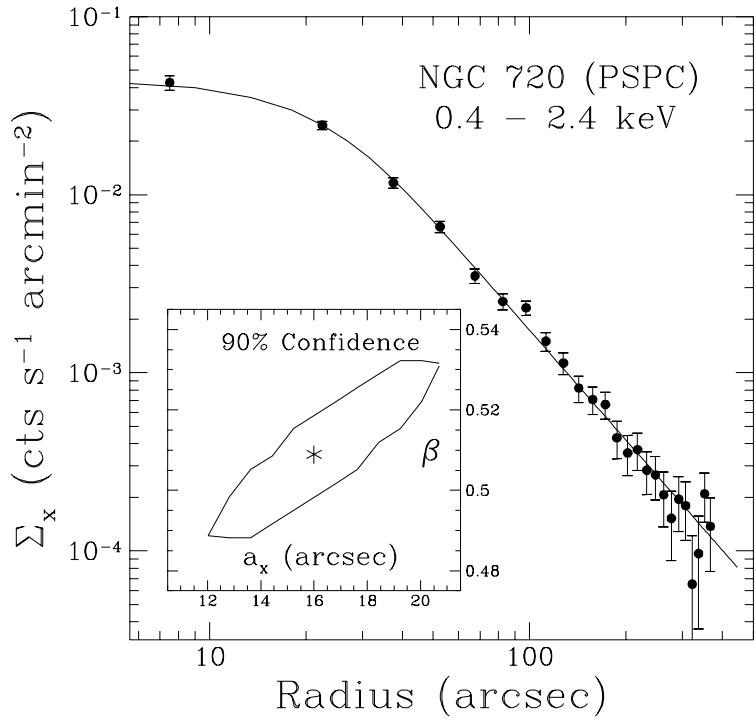

Fig. 3.—
The azimuthally-averaged radial profile of the reduced image, the best-fit King model, and the
90% confidence estimates of the fitted King parameters.

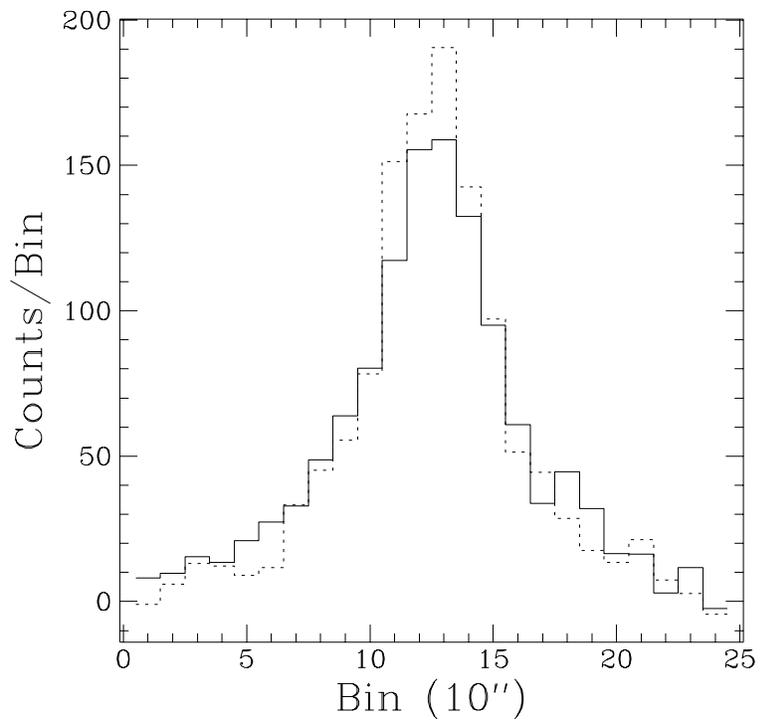

Fig. 4.—
One-dimensional projections of the image in a 240″ box along the major axis (solid) and the minor axis (dotted).

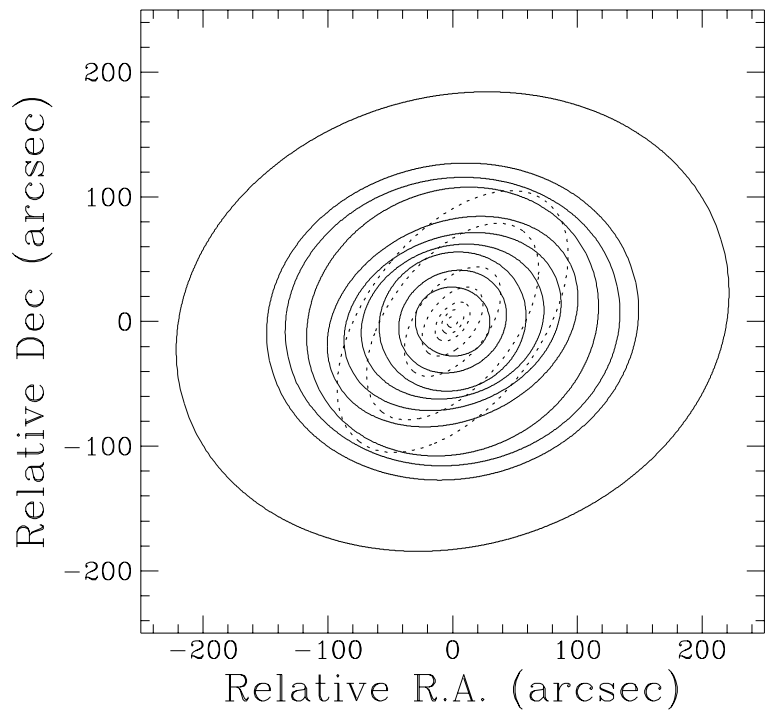

Fig. 5.—

X-ray contours (solid) having $\epsilon = \epsilon_M$ computed for the circular aperture (Table 3, column 3) and the $R$-band isophotes (dotted) from Peletier et al. (1989). The X-ray contours are separated by factors of $\sim 1.2 - 1.7$ in intensity and the optical contours are separated by 1 mag arcsec$^{-2}$.

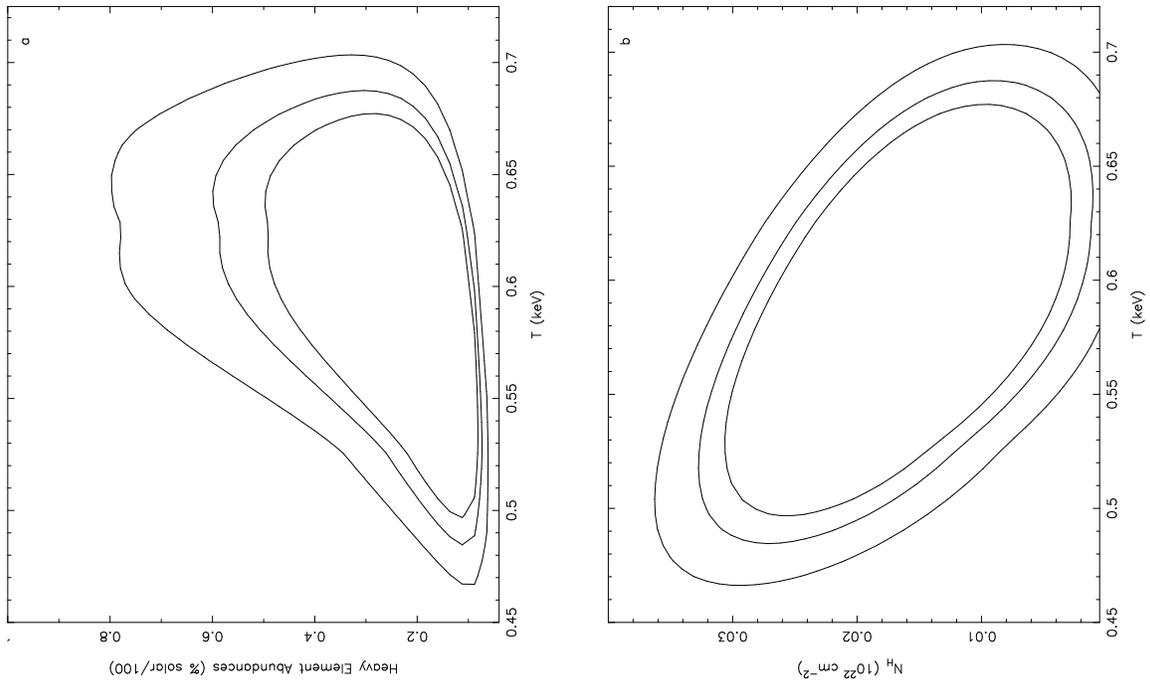

Fig. 6.—
90%, 95%, and 99% confidence contours $(0'' - 400''; 0.2 - 2.4 \text{ keV})$ for (a) abundances vs. $T$ and
(b) Hydrogen column density vs. $T$.

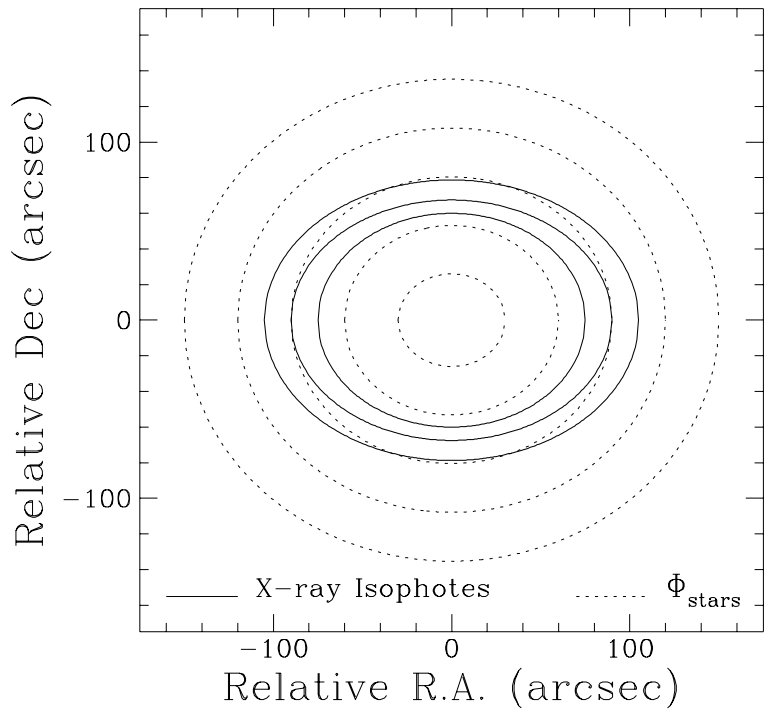

Fig. 7.—
Gravitational potential projected onto the plane of the sky (dotted) generated by mass distributed like the stars; the ellipticities are those of 2-D $\epsilon_{pot}$ in Table 6. For comparison, the most distant X-ray isophotes whose shapes are very accurately determined are also plotted as perfect ellipses (solid). The relative position angle offset of the X-ray and optical isophotes is suppressed.

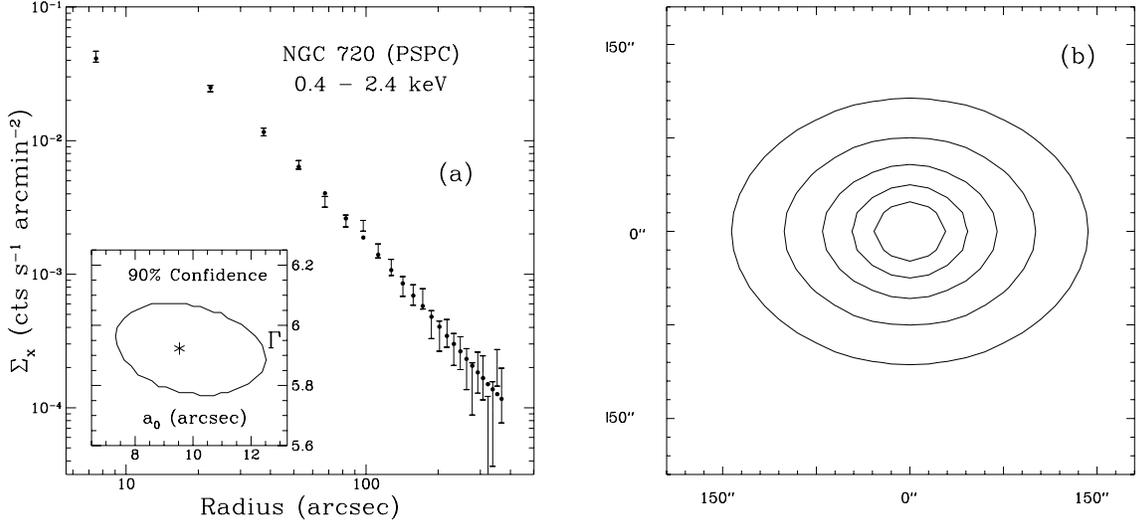

Fig. 8.—

(a) Radial profile of a typical model (filled circles) consistent with the data (error bars); the model displayed is oblate and has $\rho_{tot} \sim r^{-2}$, $a_3 = 450''$, $\epsilon_{tot} = 0.60$, $a_0 = 9.5''$, $\Gamma = 5.92$, and $\chi^2_{min} = 14.8$. The 90% confidence contour and the best-fit values are displayed in the inset.

(b) X-ray isophotes for the best-fit model ($375'' \times 375''$) separated by a factor of 2 in intensity.

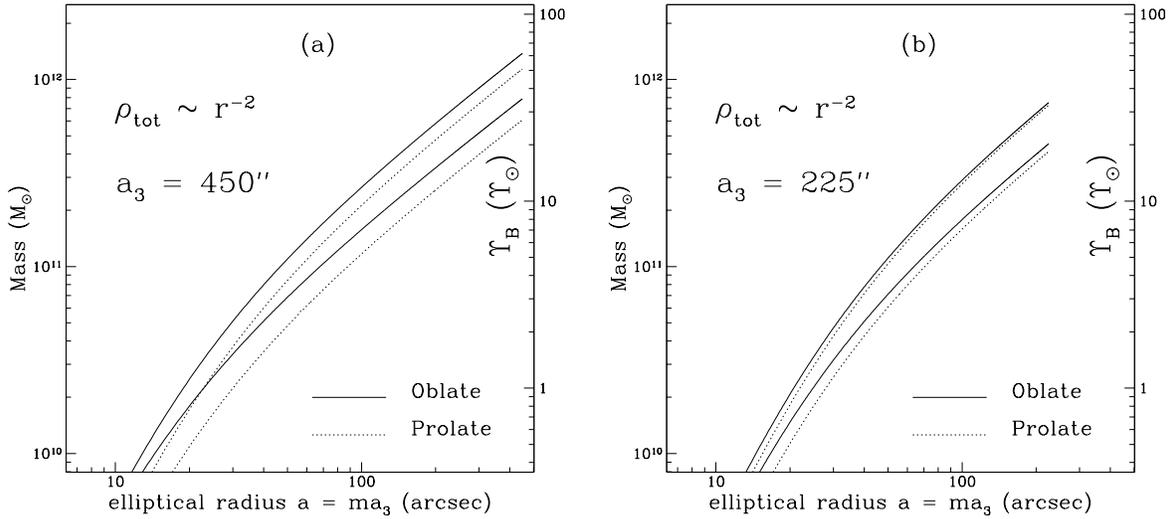

Fig. 9.—

Upper and lower solid (dashed) curves show 90% confidence limits of the integrated mass as a function of elliptical radius for the galaxy modeled as a single oblate (prolate) ellipsoid having $\rho_{tot} \sim r^{-2}$ and semi-major axis (a) $a_3 = 450''$ and (b) $a_3 = 225''$

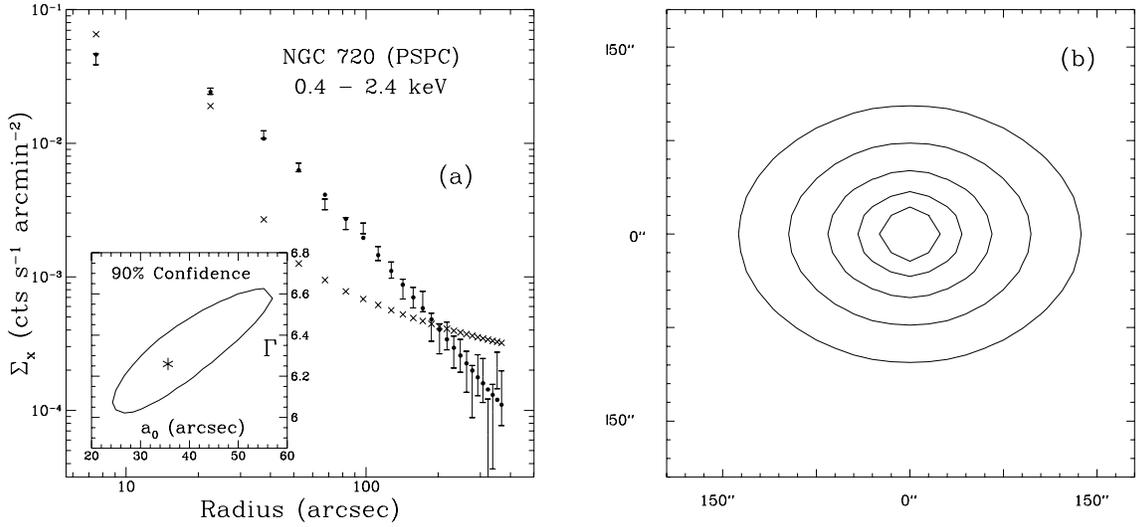

Fig. 10.—

(a) Radial profile of (1) the model without dark matter (crosses) and (2) a typical model (filled circles) consistent with the data (error bars): $\epsilon_{DM} = 0.60$, $a_3 = 450''$, $M_{DM} = 10M_{stars}$, $M_{gas} = M_{stars}/50$, and $M_{stars} = \Upsilon_B L_B = 1.6 \times 10^{11} M_{\odot}$, where $\Upsilon_B \sim 7\Upsilon_{\odot}$ is the $B$-band mass-to-light ratio of the stellar matter in solar units. The 90% confidence contour and the best-fit values are displayed in the inset.

(b) X-ray isophotes for the best-fit model ($375'' \times 375''$) separated by a factor of 2 in intensity.

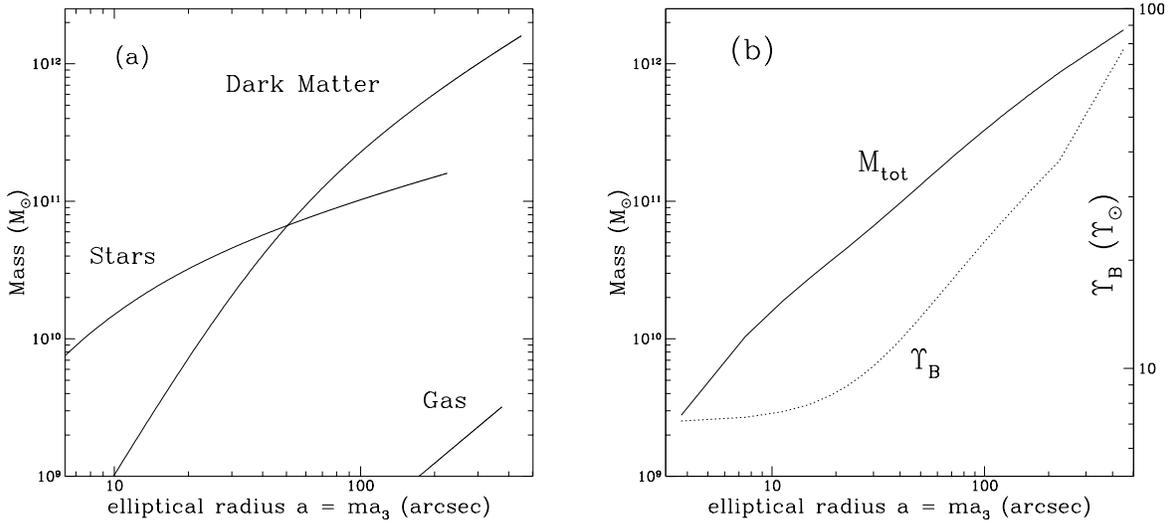

Fig. 11.—

(a) Integrated mass as a function of elliptical radius for the dark matter, stars, and gas corresponding to the model of Figure 10. Interior to $\sim 50''$ (which is the radius enclosing half the light) $M_{stars}$ dominates $M_{DM}$ while the opposite is true for larger distances. The self-gravitation of the gas is not important anywhere in the galaxy.

(b) Here we show the total mass for this model and the corresponding $\Upsilon_B$ as a function of ellipsoidal radius. $\Upsilon_B$ is very nearly constant inside of $10''$ but increases substantially for $a > 20''$. Note that the "kink" in $\Upsilon_B$ at $a = 225''$ occurs where we assign the discrete edge to the stellar mass; this can be made smooth by adding an exponential cutoff.

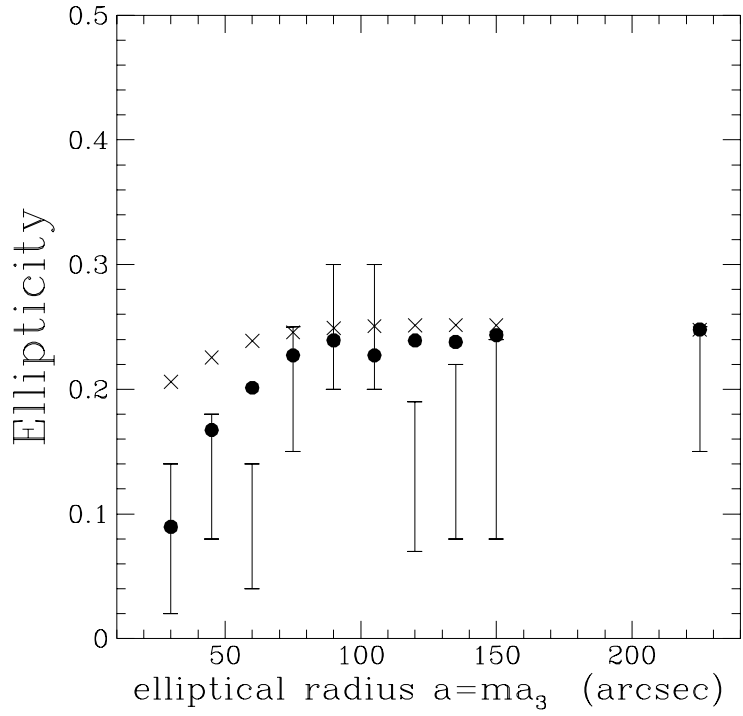

Fig. 12.—
Shown are the results of computing $\epsilon_M$ in an elliptical aperture of semi-major axis $a$ (filled circles) and the actual ellipticity of the isophote at $a$ (crosses) for the total mass model in Figure 8; also displayed is $\epsilon_M$ computed from the data listed in Table 3 (error bars).

Fig. 13.—
Shown are the results of the projection (edge-on) of the pseudo oblate spheroids discussed in
Appendix B. for the three regimes (a), (b), and (c) of interest. We plot the $\epsilon$ of (1) the three
dimensional surfaces (i.e. $\epsilon(r)$, small dashes), (2) the projection of $(a_0^2 + \xi^2)^{-2}$ (big dashes), (3) the
projection of $\log(a_0^2 + \xi^2)$ (solid line), and (4) the difference in ellipticity of (2) and (3) (dot-dash).

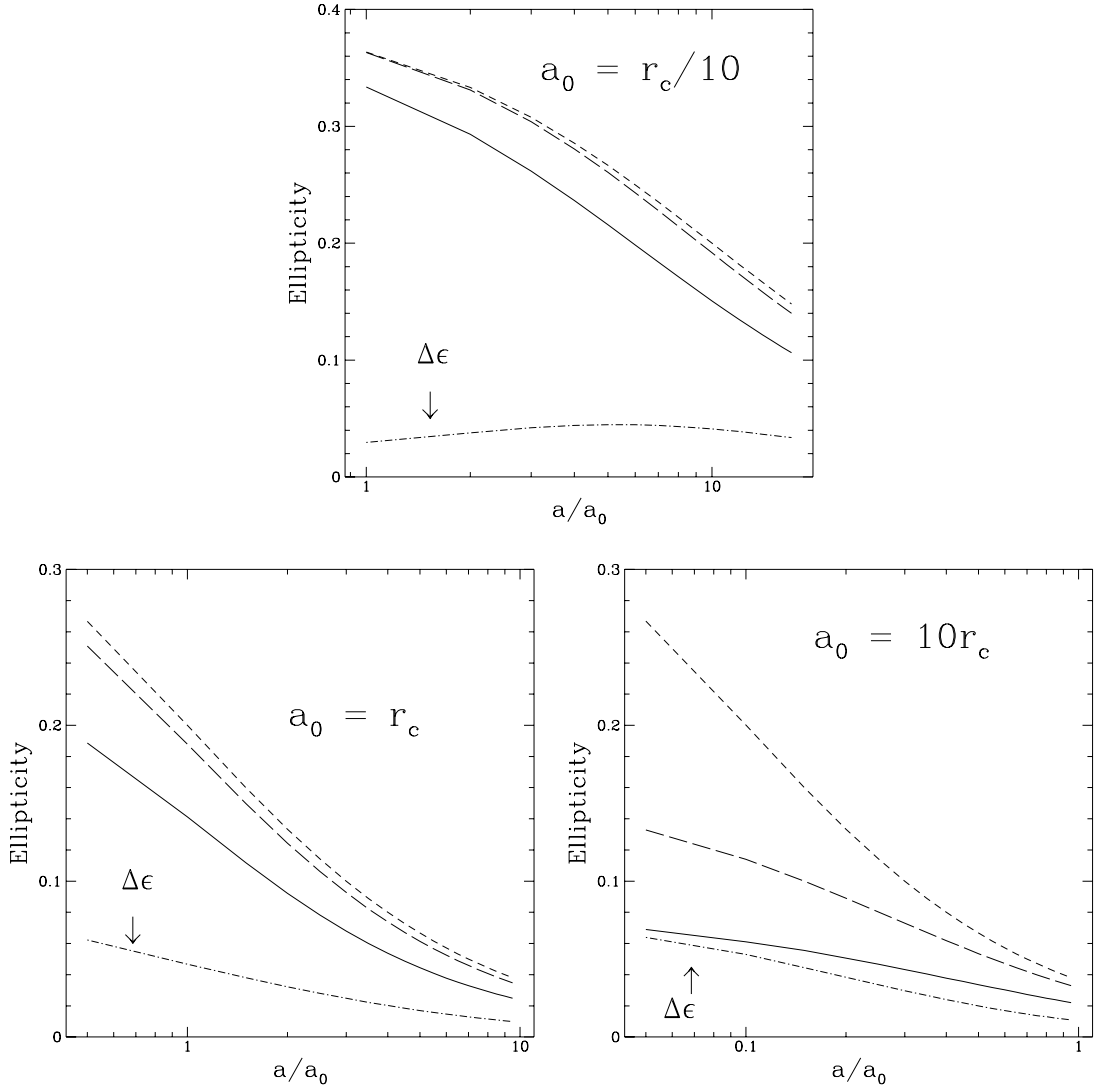

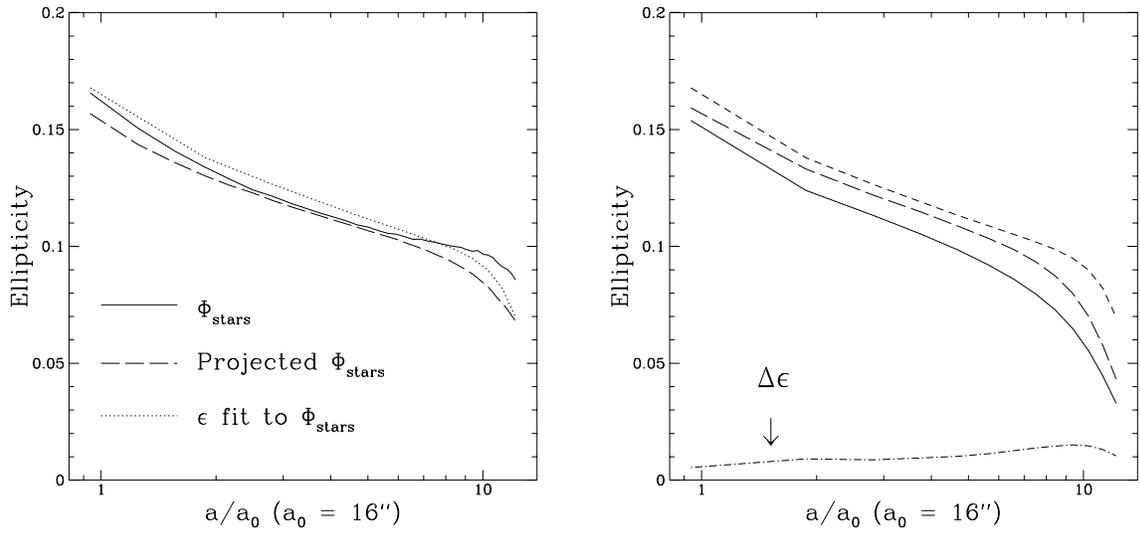

Fig. 14.—

(a) Major-axis ellipticity profile of model for $\Phi_{stars}$ (solid line) and its edge-on projection (dashes). The fit to the three-dimensional ellipticity that is used for the ensuing pseudo spheroid comparison is given by the dotted line.

(b) Results of the edge-on projections of the pseudo oblate spheroids discussed in Appendix B. corresponding to the gas emissivity (big dashes) and stellar gravitational potential of NGC 720 (solid line). The small dashes represent the approximation to the three dimensional ellipticity (see (a)) of the stellar potential and $\Delta\epsilon$ is the ellipticity difference of the projections.